\documentclass[graybox]{svmult}

\usepackage{mathptmx}       
\usepackage{helvet}         
\usepackage{courier}        
\usepackage{type1cm}        
\usepackage{makeidx}         
\usepackage{graphicx}        
\usepackage{multicol}        
\usepackage[bottom]{footmisc}

\makeindex             

\def\lya{Lyman-$\alpha$~}

\def\centering{ }

\def\citep{\cite}
\def\citet{\cite}

\def\ga{>}

\def\la{<}

\def\lesssim{<}

\def\beq{\begin{equation}}
\def\eeq{\end{equation}}
\def\ba{\begin{eqnarray}}
\def\ee{\end{equation}}

\def\beqa{\begin{eqnarray}}
\def\eeqa{\end{eqnarray}}

\newcommand{\bk}{{\bf k}}

\def\VEV#1{\left\langle #1\right\rangle} 

\newcommand{\Mpcinv}{\mbox{ Mpc$^{-1}$}}

\newcommand{\sqm}{\mbox{ m$^2$}}

\newcommand{\yrinv}{\mbox{ yr$^{-1}$}}
\newcommand{\ergsec}{\mbox{erg s$^{-1}$}}
\newcommand{\hr}{\mbox{ hr}}
\newcommand{\Hz}{\mbox{ Hz}}

\newcommand{\MHz}{\mbox{ MHz}}

\newcommand{\Jnunits}{\mbox{ cm$^{-2}$ s$^{-1}$ Hz$^{-1}$ sr$^{-1}$}}
\newcommand{\recunits}{\mbox{ cm$^{3}$ s$^{-1}$}}
\newcommand{\kel}{\mbox{ K}}
\newcommand{\mkel}{\mbox{ mK}}

\begin{document}

\title*{The 21-cm Line as a Probe of Reionization}
\author{Steven R. Furlanetto}
\institute{Steven R. Furlanetto \at UCLA Department of Physics \& Astronomy, Los Angeles, CA 90095, USA; \email{sfurlane@astro.ucla.edu}}
%
%
\maketitle

\abstract{One of the most exciting probes of the early phases of structure formation and reionization is the ``spin-flip" line of neutral hydrogen, with a rest wavelength of 21 cm. This chapter introduces the physics of this transition and the astrophysical parameters upon which it depends, including discussions of the radiation fields that permeate the intergalactic medium that fix the brightness of this transition. We describe the critical points in the evolution of the 21-cm background and focus on the sky-averaged brightness and the power spectrum as representative measurements. Finally, we include a discussion of observations and the challenges they face in the near future.}

\section{Introduction} \label{sec:intro}

Although the \lya line and the CMB are extremely powerful probes of reionization, they suffer from several shortcomings.  Most importantly, the Gunn-Peterson optical depth is enormous, so that even a very small fraction of neutral hydrogen ($\ga 10^{-3}$) saturates the IGM absorption. The \lya line is therefore difficult to interpret during the middle and early stages of reionization.  On the other hand, the CMB probes are integrated measurements along the line of sight, offering no (direct) discriminatory power between events at different redshifts.

These problems can be avoided by observing the \emph{spin-flip} or hyperfine line of neutral hydrogen, which is driven by the magnetic interactions of the proton and electron -- though of course such a strategy introduces a new set of problems.   Hendrik van de Hulst first predicted the existence of this transition \cite{vdhulst45} (after a suggestion by Jan Oort), and Harold Ewen and Ed Purcell first observed it from our own Galaxy in 1951 \cite{ewen51}.  This transition is extremely weak, making the effective IGM optical depth only $\sim 1$\%. While the signal is therefore very faint, the  neutral IGM is accessible over the entire epoch of reionization. Moreover, the transition energy is so low that it provides a sensitive calorimeter of the diffuse IGM, and -- as a low-frequency radio transition -- it can be observed across the entire sky and be used to ``slice" the universe in the radial direction, thanks to the cosmological redshift.  With such three-dimensional observations, the 21-cm line allows \emph{tomography} of the neutral IGM, potentially providing a map of $\ga 90\%$ of the Universe's baryonic matter during the Dark Ages and cosmic dawn \cite{madau97}. As we shall see, however, there are enormous obstacles to fully utilizing this signal.

\begin{figure}[!t]
\includegraphics[scale=0.4]{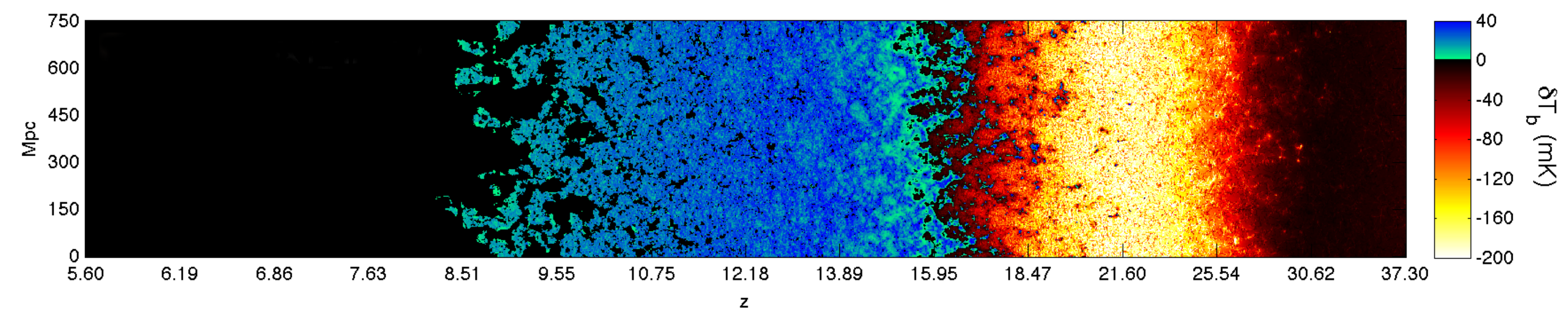}
\vskip -0in
\caption{Time evolution of the expected 21 cm signal from a semi-numeric simulation 750~Mpc on a side, spanning the period before the first stars formed (at right) through the end of reionization (at left).Galaxy parameters are similar to those of present-day galaxies. Coloration indicates the strength of the 21 cm brightness as it transitions from absorption (red) to emission (blue) and finally disappears (black) due to reionization. From \cite{valdes13}. }
\label{fig:21cm-overview}
\end{figure}

Figure~\ref{fig:21cm-overview} shows an overview of the expected spin-flip signal (taken from \cite{valdes13}).  It can be observed in two fundamental ways.  The first is the sky-averaged, or monopole, brightness, which measures the average properties of the H~I as a function of redshift.  The bottom panel shows this signal relative to the CMB.  The top panel shows the fluctuations in the 21-cm signal,\footnote{Following convention, we will often refer to the signal as ``21-cm radiation," although of course the \emph{observed} wavelengths are larger by a factor of $(1+z)$.} which arise from the discrete, clustered luminous sources.  We will discuss these probes in detail later in this chapter after introducing the physics of the 21-cm line.  Finally, we conclude with a discussion of this signal's observational prospects.  We refer the reader to several recent reviews for more information \cite{furl06-review,morales10,pritchard12,loeb13}.

\section{Fundamentals of the 21-cm Line} \label{rt-21cm}

We begin with an introduction to the atomic physics and radiative transfer of the spin-flip transition.  The radiative transfer equation for the specific intensity $I_\nu$ of a line reads
\begin{equation}
{dI_\nu\over d\ell}={\phi(\nu) h\nu\over 4\pi}\left[n_1 A_{10} -
\left(n_0 B_{01} -n_1 B_{10}\right)I_\nu\right],
\label{rad}
\end{equation}
where $d\ell$ is a proper path length, $\phi(\nu)$ is the line profile function normalized by $\int \phi(\nu)d\nu=1$, subscripts 0 and 1 denote the lower and upper atomic levels, $n_i$ are the number density of atoms in these levels, and $A_{ij}$ and $B_{ij}$ are the Einstein coefficients for the transition (with $i$ and $j$ the initial and final states, respectively). In our case, the line frequency is $\nu_{21} = 1420.4057$~MHz, corresponding to a wavelength of $\lambda_{21}=21.1061$ cm. For the 21-cm transition, $A_{10}=2.85\times 10^{-15} \ {\rm s^{-1}}$ and $g_1/g_0=3$.

The relative populations of the two spin states define the \emph{spin temperature}, $T_S$, through the relation,
\begin{equation}
\left({n_1\over n_0}\right)=\left({g_1 \over g_0}\right)
\exp\left\{ {-T_*\over T_S}\right\}, 
\end{equation}
where $g_i$ are the spin degeneracy factor of each state and $T_* \equiv E_{10} /k_B=68$~mK is equivalent to the transition energy $E_{10}$. We will always find that $T_\star$ is much smaller than the other relevant temperatures ($T_S$ and the CMB temperature $T_\gamma$), so all exponentials like this one can be expanded to leading order. Moreover, this also implies that $\sim 3/4$ of atoms are in the upper state at any time, making stimulated emission an important process.

Following convention, we will quantify $I_{\nu}$ by  the equivalent {\it brightness temperature}, $T_b(\nu)$, required of a blackbody radiator (with spectrum $B_{\nu}$) such that $I_{\nu}=B_{\nu}(T_b)$. At the low frequencies of interest to us, $T_b(\nu)\approx I_{\nu} \, c^2/2k_B{\nu}^2$ according to the Rayleigh-Jeans limit.
Then the equation of radiative transfer  along a line of sight through a cloud of uniform excitation temperature $T_S$ becomes \cite{madau97}
\begin{equation}
T_b'(\nu) = T_{S}(1-e^{-\tau_{\nu}})+T_R'(\nu)e^{-\tau_{\nu}}
\label{eq:rad_trans}
\end{equation}
where the {\it optical depth}  $\tau_\nu \equiv \int d s \, \alpha_{\nu}$ is the integral of the absorption coefficient ($\alpha_{\nu}$)  along the ray through the cloud, $T_R'$ is the brightness of the background radiation field incident on the cloud along the ray, and $s$ is the proper distance. Because of the cosmological redshift, the emergent brightness $T_b'(\nu_0)$ measured in a cloud's comoving frame at redshift $z$ creates an apparent brightness at the Earth of $T_b(\nu) = T_b'(\nu_0)/(1+z)$, where the observed frequency is $\nu=\nu_0/(1+z)$. Henceforth we will work in terms of these observed quantities. The absorption coefficient is determined from the Einstein coefficients via
\beq
\alpha = \phi(\nu) {h \nu \over 4 \pi} (n_0 B_{01} - n_1 B_{10}),
\eeq
where the $B_{ij}$ can be derived from the $A_{10}$ given above using the standard Einstein relations.  

In an expanding Universe with a local hydrogen number density $n_{\rm H}$ and with a velocity gradient along the line of sight of $dv_\parallel/dr_\parallel$, the 21-cm optical depth can be derived just like the Gunn-Peterson optical depth \cite{field59-obs}. Writing $\phi(\nu)\sim 1/(\Delta \nu)$, we obtain \cite{madau97}
\begin{eqnarray}
\tau_{10} & = & \frac{3}{32 \pi} \, \frac{h c^3 A_{10}}{k_B T_S \nu_{10}^2} \, \frac{x_{\rm HI} n_{\rm H}}{(1+z) \, (d v_\parallel/d r_\parallel)}  \label{eq:optdepthcosmo} \\
 & \approx  & 0.0092 \, (1+\delta) \, (1+z)^{3/2}\, \frac{x_{\rm HI}}{T_S} \, \left[ \frac{H(z)/(1+z)}{d v_\parallel/d r_\parallel} \right],
\label{optdepthcosmo-approx}
\end{eqnarray}
In the second part we express $T_S$ in Kelvins and have scaled to the mean IGM density at $z$ and to the Hubble flow (so that $\Delta I_\nu \propto \Delta \ell \phi(\nu)\nu= \vert cdt/dz\vert [\nu dz/d\nu] =c/H$).

In practice, the background radiation source is usually the CMB, so $T_R' = T_{\gamma}(z)$, and we are observing the contrast between high-redshift hydrogen clouds and the CMB.   With $\tau_\nu \ll 1$,
\begin{eqnarray}
& T_b(\nu) & \approx \frac{T_S-T_{\gamma}(z)}{1+z}\;\tau_{\nu_0} 
\label{eq:dtbone} \\
& \approx & 9\;x_{\rm HI}(1+\delta) \, (1+z)^{1/2}\, \left[1-\frac{T_{\gamma}(z)}{T_S}\right] \, \left[ \frac{H(z)/(1+z)}{d v_\parallel/d r_\parallel} \right] \mkel.
\label{eq:dtb}
\end{eqnarray}
Here $T_b < 0$ if $T_S < T_{\gamma}$, yielding an absorption signal, while $T_b > 0$ otherwise, yielding emission. Both regimes are important for the high-$z$ Universe, though the consensus is currently that emission will dominate during the reionization era. In that case, $\delta T_b$ saturates if $T_S \gg T_{\gamma}$ (though this is not true in the absorption regime).  

\subsection{The Spin Temperature} \label{spin-temp}

Three processes compete to fix $T_S$ \cite{wouthuysen52, field58, field59-ts}: {\it (i)} interactions with CMB photons; {\it (ii)} particle collisions; and
{\it (iii)} scattering of UV photons.  The CMB very rapidly drives the spin states toward thermal equilibrium with $T_S=T_{\gamma}$. However, the other two processes break this coupling. We let $C_{10}$ and $P_{10}$ be the de-excitation rates (per atom) from collisions and UV scattering, respectively (with corresponding excitation rates $C_{01}$ and $P_{01}$).  In equilibrium,
\begin{equation}
n_1 \left( C_{10} + P_{10} + A_{10} + B_{10} I_{\rm CMB} \right) = n_0 \left( C_{01} + P_{01} + B_{01} I_{\rm CMB} \right),
\label{eq:detbal}
\end{equation}
where $I_{\rm CMB}$ is the specific intensity of CMB photons.  In the Rayleigh-Jeans limit, equation (\ref{eq:detbal}) becomes
\begin{equation}
T_S^{-1} = \frac{T_\gamma^{-1} + x_c T_K^{-1} + x_\alpha T_c^{-1}}{1 + x_c + x_\alpha},
\label{eq:xdefn}
\end{equation}
where $x_c$ and $x_\alpha$ are coupling coefficients for collisions and UV scattering, respectively, and $T_K$ is the gas kinetic temperature.  Here we have used the principle of detailed balance through the relation
\begin{equation}
\frac{C_{01}}{C_{10}} = \frac{g_1}{g_0} e^{-T_\star/T_K} \approx 3 \left( 1 - \frac{T_\star}{T_K} \right).
\label{eq:c01db}
\end{equation}
We have also \emph{defined} the effective color temperature of the UV radiation field $T_c$ via
\begin{equation}
\frac{P_{01}}{P_{10}} \equiv 3 \left( 1 - \frac{T_\star}{T_c} \right).
\label{eq:tcolor}
\end{equation}

In the limit in which $T_c \rightarrow T_K$ (a reasonable approximation in most situations of interest), equation~(\ref{eq:xdefn}) may be written as
\begin{equation}
1 - \frac{T_\gamma}{T_S} = \frac{x_c + x_\alpha}{1 + x_c + x_\alpha} \, \left( 1 - \frac{T_\gamma}{T_K} \right).
\label{eq:xdefn-tfac}
\end{equation}
Thus, particle collisions and photons both tend to drive $T_S \rightarrow T_K$; to understand the signal, we must understand how strong these coupling processes are and the IGM's thermal history.

The collisional coupling coefficient for collisions with species $i$ is
\begin{equation}
x_c^i \equiv  \frac{C_{10}^i}{A_{10}} \, \frac{T_\star}{T_\gamma} = \frac{n_i \, \kappa_{10}^i}{A_{10}} \, \frac{T_\star}{T_\gamma},
\label{eq:xcdefn}
\end{equation}
where $\kappa_{10}^i$ is the rate coefficient for spin de-excitation in collisions with that species (with units of $\recunits$).  The total $x_c$ is the sum over all species $i$, which are generally dominated by collisions with (1) other hydrogen atoms \cite{zygelman05} and (2) free electrons \cite{furl07-electron}.  Although the atomic cross-section is small, in the unperturbed IGM collisions between neutral hydrogen atoms nearly always dominate these rates because the ionized fraction is small.  Free electrons can be important in partially ionized gas.

Crucially, the collisional coupling is quite weak in a nearly neutral,
cold medium.  Thus, the overall density must be large in order for this
process to effectively fix $T_S$. A convenient estimate of their
importance is the critical overdensity, $\delta_{\rm coll}$, at which
$x_c=1$ for H--H collisions:
\begin{equation}
1 + \delta_{\rm coll} = 0.99 \, \left[ \frac{\kappa_{10}(88
    \kel)}{\kappa_{10}(T_K)} \right] \, \left( \frac{0.023}{\Omega_b
    h^2} \right) \, \left( \frac{70}{1+z} \right)^2,
\label{eq:dcoll}
\end{equation}
where we have inserted the expected temperature at $1+z=70$.  In the standard picture, at
redshifts $z \la 70$, $x_c \ll 1$ and $T_S \rightarrow T_{\gamma}$; by $z \sim 30$
the IGM essentially becomes invisible.  

\begin{figure}[!t]
\sidecaption
\includegraphics[scale=0.3]{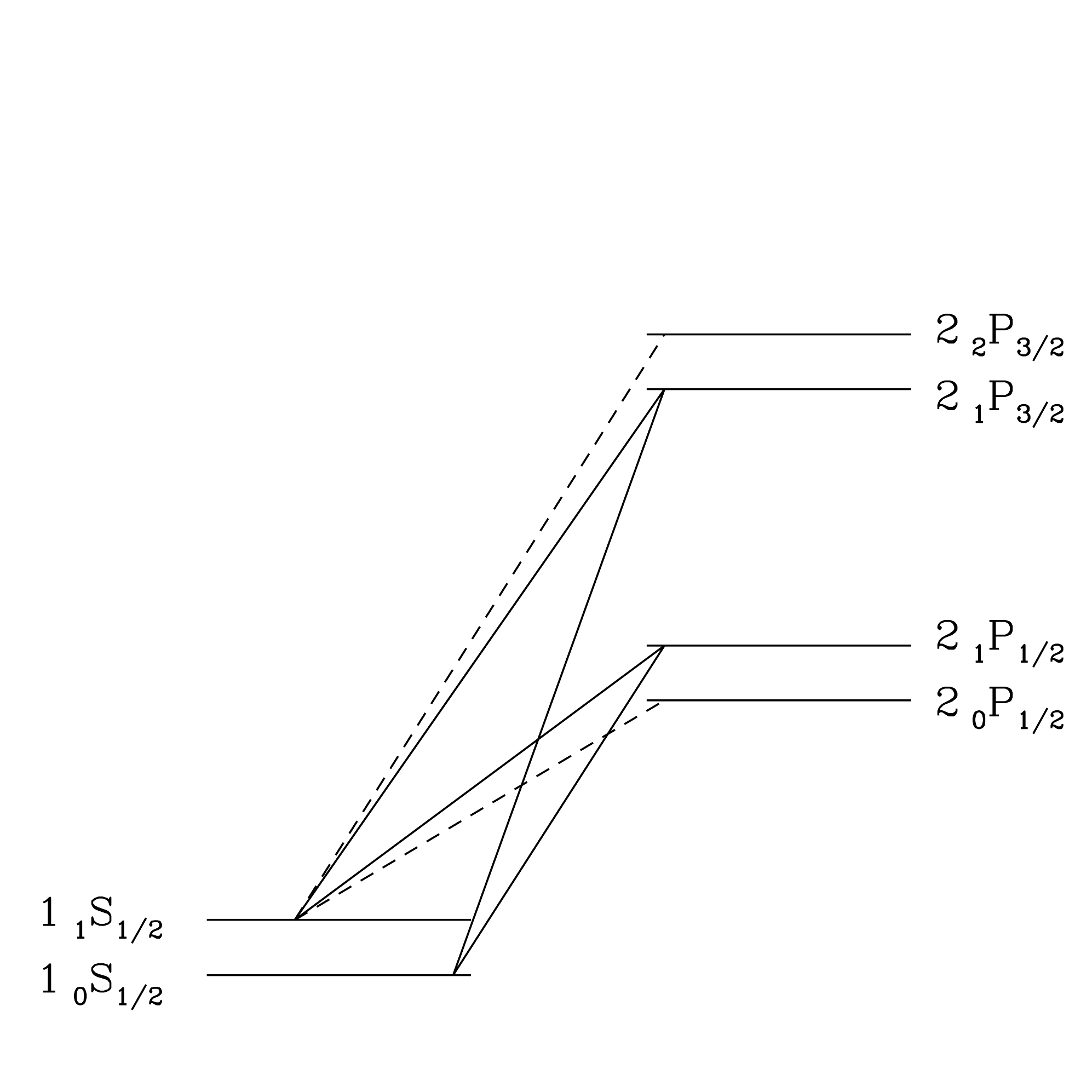}
\caption{Level diagram illustrating the Wouthuysen-Field effect.  We
show the hyperfine splittings of the $1S$ and $2P$ levels.  The solid
lines label transitions that mix the ground state hyperfine levels,
while the dashed lines label complementary allowed transitions that do not
participate in mixing.  From \cite{pritchard06}. }
\label{fig:wf}
\end{figure}

We therefore require a different process to break the coupling to
the CMB during the era of galaxy formation.  The \emph{Wouthuysen-Field mechanism} \cite{wouthuysen52,field58,hirata06} provides just such an effect. Figure~\ref{fig:wf} shows the process, where we have drawn the hyperfine
sub-levels of the $1S$ and $2P$ states of HI.  Suppose a hydrogen atom
in the hyperfine singlet state absorbs a \lya photon.  The electric
dipole selection rules allow the electron to jump to either of the central $2P$
states.  However, these states can decay to the upper hyperfine level, changing the hyperfine populations through the
absorption and spontaneous re-emission of a \lya photon (or any other
Lyman-series photon, though those only contribute significantly if they produce \lya as a cascade product \cite{hirata06,pritchard06}).

The Wouthuysen-Field coupling strength depends on the total rate (per atom) at which \lya photons scatter through the gas, 
\begin{equation}
P_\alpha = 4 \pi \sigma_0 \int d \nu \, J_\nu(\nu) \phi_\alpha(\nu),
\label{eq:palpha}
\end{equation}
where $\sigma_\nu \equiv \sigma_0 \phi_\alpha(\nu)$ is the local absorption cross section, $\sigma_0 \equiv (\pi \, e^2/m_e \, c) f_{\alpha}$, $f_\alpha=0.4162$ is the oscillator strength of the \lya
transition, $\phi_\alpha(\nu)$ is the \lya absorption profile, and
$J_\nu$ is the angle-averaged specific intensity of the background
radiation field, in units of photons cm$^{-2}$ Hz$^{-1}$ s$^{-1}$ sr$^{-1}$ here.

Not all of these scattered photons contribute to hyperfine level changes, however, so we must relate $P_\alpha$ to the indirect excitation and de-excitation rates $P_{01}$ and $P_{10}$ \cite{meiksin00}. To do so, we first relabel the $1S$ and $2P$ hyperfine levels a--f, in order of increasing energy, and let $A_{ij}$ and $B_{ij}$ be the spontaneous emission and absorption coefficients for transitions between these levels.  We write the background flux at the frequency corresponding to the $i \rightarrow j$ transition as $J_{ij}$.  Then
\begin{equation}
P_{01} \propto B_{\rm ad} J_{\rm ad} \frac{A_{\rm db}}{A_{\rm da} + A_{\rm db}} + B_{\rm ae} J_{\rm ae} \frac{A_{\rm eb}}{A_{\rm ea} + A_{\rm eb}}.\label{eq:psum}
\end{equation}
The first term contains the probability for an a$\rightarrow$d transition ($B_{\rm ad} J_{\rm ad}$), multiplied by the probability for the subsequent decay to terminate in state b; the second term is the same for transitions to and from state e.  Next we need to relate the individual $A_{ij}$ to $A_\alpha = 6.25 \times 10^8 \Hz$, the total \lya spontaneous emission rate (averaged over all the hyperfine sublevels).  This can be accomplished using a sum rule stating that the sum of decay intensities ($g_i A_{ij}$) for transitions from a given $nFJ$ to all the $n' J'$ levels (summed over $F'$) is proportional to $2F+1$ \cite{bethe57}. The relative strengths of the permitted
transitions are then $(1,\,1,\,2,\,2,\,1,\,5)$, where we have ordered the lines (bc, ad, bd, ae, be, bf).  Assuming that the background radiation field is constant across the individual hyperfine lines, we find $P_{10} = (4/27) P_\alpha$.

The coupling coefficient $x_\alpha$ may then be written
\begin{equation}
x_\alpha = \frac{4 P_\alpha}{27 A_{10}} \, \frac{T_\star}{T_{\gamma}} = S_\alpha \frac{J_\alpha}{J_\nu^c},
\label{eq:xalpha}
\end{equation}
where in the second equality we evaluate $J_\nu$ at line center and set $J_\nu^c \equiv 1.165 \times 10^{-10} [(1+z)/20] \Jnunits$.  We include here  a correction factor $S_\alpha < 1$ that accounts for variations in the intensity near the line center that is typically of order unity \cite{chen04, hirata06, furl06-lyheat}.  (Intuitively, a flat input spectrum develops an absorption feature because of the increased scattering rate near the \lya resonance. Photons continually lose energy by redshifting, but they also lose energy through recoil whenever they scatter.) This coupling threshold for $x_\alpha = S_\alpha$ can also be written in terms of the number of \lya photons per hydrogen atom in the Universe, which we denote $\tilde{J}_\nu^c = 0.0767 \, [(1+z)/20]^{-2}$.  This threshold is relatively easy to achieve in practice.

The remaining challenge is to compute $T_c$, the effective temperature of the UV radiation field. A simple argument shows that $T_c \approx T_K$ \cite{wouthuysen52, field59-res}: so long as the medium is extremely optically thick, the enormous number of \lya scatterings must bring the \lya profile to a blackbody of temperature $T_K$ near the line center.  This condition is easily fulfilled in the high-redshift IGM, where $\tau_\alpha \gg 1$.  In detail, atomic recoils during scattering tilt the spectrum to the red and are primarily responsible for establishing this equilibrium.  

\section{The Brightness Temperature of the 21-cm Background} \label{tb-hist}

Next we consider the astrophysical processes that drive the 21-cm background.  In general terms, three important radiation backgrounds affect the signal: (1) the metagalactic field near the \lya resonance, which determines the strength of the Wouthuysen-Field effect; (2) the X-ray background, which determines the amount of IGM heating; and (3) the ionizing background, which eventually (nearly) eliminates the signal at the completion of reionization.  We will discuss each of these in turn in this section.

\subsection{The Lyman-$\alpha$ Background} \label{wf-bkgd}

After $z \sim 30$, when collisional coupling becomes unimportant, the spin temperature is determined by the scattering of \lya photons.  In practice, the relevant photons do not start at the Lyman-$\alpha$ wavelength, because those redshift out of resonance very soon after they are created and do not contribute to the coupling except very near their sources.  Instead, the important photons begin in the ultraviolet and redshift into a Lyman-series line, possibly cascading down to a \lya photon.

To compute $J_\alpha$, we therefore begin with the proper ultraviolet emissivity at a frequency $\nu$, $\epsilon(\nu,z)$.  For the purposes of a simple global estimate, we will consider the limit in which this emissivity is nearly uniform. Then 
\begin{eqnarray}
J_\alpha(z) & = & \sum_{n=2}^{n_{\rm max}} J_\alpha^{(n)}(z) \nonumber
\\
& = & {c \over 4 \pi} \sum_{n=2}^{n_{\rm max}} f_{\rm rec}(n) \int_z^{z_{\rm max}(n)} 
d z' \, \left| {dt \over dz'} \right| \left( {1+z \over 1+z'}\right)^3{4 \pi} \, \frac{c}{H(z')} \,
\epsilon(\nu_n',z'),
\label{eq:jalpha}
\end{eqnarray}
where $\nu_n'$ is the frequency at redshift $z'$ that redshifts into the Lyman-$n$ resonance at redshift $z$,  $z_{\rm max}(n)$ is the largest redshift from which a photon can redshift into the Lyman-$n$ resonance, and $f_{\rm rec}(n)$ is the fraction of Ly$n$ photons that produce a \lya photon as part of their cascade.  The sum must be truncated at some large $n_{\rm max}$ that is determined by the typical size of ionized regions around the sources, but the result is not sensitive to the precise cutoff value.

To estimate the background, we need to understand the sources of \lya photons, most likely star-forming galaxies (though X-rays can also produce them as fast electrons scatter through the IGM and collisionally excite hydrogen atoms). In the simplest model, in which the star formation rate traces the rate at which matter collapses into galaxies, the comoving emissivity at frequency $\nu$ is
\begin{equation}
\epsilon(\nu,z) = f_\star \, {\rho_b \over m_p} \, \epsilon_{{\rm L}n}(\nu) \,
\frac{d f_{\rm coll}}{d t},
\label{eq:sfemiss}
\end{equation}
where $f_\star$ is the fraction of baryonic material converted to stars, $\rho_b$ is the average baryon density, $\epsilon_{{\rm L}n}(\nu)$ is the number of photons produced in the frequency interval $\nu \pm d \nu/2$ per baryon incorporated into stars, and $f_{\rm coll}$ is the fraction of matter inside star-forming halos.  Although real spectra are rather complicated, a useful quantity is the total number $N_\alpha$ of photons per baryon in the interval 10.2--13.6~eV.  For low-metallicity Pop II stars and very massive Pop III stars, this is $N_\alpha=9690$ and $N_\alpha=4800$, respectively \cite{barkana05-ts}.

The Lyman-$\alpha$ background at any given point in space samples a very large background of sources: the effective ``horizon" within which a given galaxy is visible is $\sim 250$~comoving Mpc \cite{ahn09, holzbauer12}. The fractional fluctuations in the Lyman-$\alpha$ background thus tend to be \emph{relatively} smaller than those in later eras, but the large absolute value of the absorption signal means that the actual level of fluctuations (as measured in mK, for example) can still be relatively large \cite{pritchard06, mesinger14}. Moreover, this horizon is comparable to the scales over which the relative baryon and dark matter velocities vary \cite{tseliakhovich10}, so those velocity features can induce much stronger variations in the Wouthuysen-Field coupling in some circumstances \cite{dalal10,visbal12}. In \S \ref{21-flucs} we will consider how variations in this background translate into fluctuations in the 21-cm signal.

\subsection{The X-Ray Background} \label{igm-heat}

The Wouthuysen-Field background couples the spin temperature to the gas kinetic temperature, so we must also compute the latter.  A number of processes may contribute to it: shock heating from structure formation \cite{furl04-sh, kuhlen06-21cm, mcquinn12}, ultraviolet photons \cite{chen04, furl06-lyheat}, and X-rays \cite{ciardi03-21cm, pritchard07}. The last is thought to dominate in nearly all cases -- whether from active galactic nuclei, supernova remnants \cite{oh01}, stellar-mass black holes \cite{oh01, furl06-glob, mirabel11, fragos13, mirocha14, fialkov14}, or hot ISM thermal emission \cite{pacucci14}.  We will consider stellar-mass black hole remnants of massive stars as a fiducial model, but any or all of these can be significant.

The simplest way to parameterize this emissivity is with the local correlation between the star formation rate (SFR) and the X-ray luminosity in the photon energy band of 0.5--8~keV \cite{mirabel11},
\begin{equation}
L_X = 3 \times 10^{39} f_X \left( \frac{{\rm SFR}}{M_\odot \yrinv} \right)
\ergsec,
\label{eq:sfrxray}
\end{equation}
where $f_X$ is an unknown renormalization factor appropriate for high redshifts.  We can only speculate as to the appropriate value at higher redshifts.  Certainly the scaling is appropriate so long as recently-formed remnants dominate, but $f_X$ will likely evolve through several factors. Qualitatively, for example: (1) a decreasing metallicity appears to increase the relative efficiency of X-ray production \cite{fragos13-model};  (2) if the IMF becomes more top-heavy, the total X-ray luminosity will increase as the abundance of stellar remnant black holes increases, but the spectra may also harden and thereby \emph{decrease} the fraction of X-ray energy that can be absorbed by the IGM \cite{fialkov14, mirocha14}; (3) smaller galaxies at high-redshifts may systematically change the column density of neutral gas that absorbs the X-rays (as required by suggestions that the escape fraction of UV photons increase toward higher redshift; \cite{haardt12, kuhlen12}); and (4) the increasing CMB energy density at high redshifts may make inverse-Compton emission from supernova remnants more important \cite{oh01}.

Once the source properties are established (or, more likely, guessed), it is straightforward to compute the evolving IGM temperature through a framework analogous to equation~{\ref{eq:jalpha}). The key difference is that X-rays are not absorbed in resonance lines but through photoionization of H or He, for which the cross section is a strong function of frequency (varying approximately as $\nu^{-3}$ near the ionization threshold). The comoving mean free path of an X-ray photon with energy $E$ through the neutral IGM is
\begin{equation}
\lambda \approx 4.9 \left( {1+z \over 15} \right)^{-2} \left( {E \over 300 \, {\rm eV}} \right)^3 \ {\rm Mpc}.
\label{eq:xray-mfp}
\end{equation}
This is much smaller than the typical path length of a photon redshifting into a Lyman resonance, so the (low-enegy) X-ray background fluctuates quite strongly, with the heating preferentially occurring near the X-ray sources \cite{pritchard07}. On the other hand, the steep energy dependence of this expression suggests that the universe is transparent to the hard X-ray background.  For hard sources (as in some models of X-ray binaries with strong local absorption; \cite{fragos13, fialkov14}), the effective value of our $f_X$ parameter may be much less than unity.

An additional complication is that X-rays deposit only a fraction of their energy as heat: they initially interact with the IGM by ionizing a neutral atom. The high-energy photoelectron then scatters through the IGM, ionizing more atoms, collisionally exciting others, and heating the gas through scattering off of other electrons.  The fraction of energy deposited in each of these processes varies with photon energy and the ambient conditions \cite{shull85, furl10-xray}, but as a rule of thumb each gets about 1/3 of the total for nearly-neutral gas.

\subsection{The Ionizing Background} \label{ion-bkgd}

The process of reionization is discussed in great detail elsewhere in this volume, so we only touch upon the major issues here.  For stellar sources of reionization, the photons are sufficiently close to the ionization threshold that they are very quickly absorbed by the neutral IGM. This creates (at least to first-order) a two-phase medium, with highly-ionized regions surrounding the sources surrounded by a sea of neutral gas. 

In that case, the key issue is the spatial distribution of these ionized regions rather than the local amplitude of the ionizing background. How big are they? How do they correlate with the underlying density field? How do they connect to each other? These issues were first attacked through analytic models \cite{furl04-bub, furl05-rec} but are now generally addressed through ``semi-numeric" simulations \cite{mesinger07, mesinger11}, which apply analytic arguments to generate the ionization field in a large slice of the universe.

Of course, the H~II regions are neither homogeneous nor perfectly ionized: dense clumps can recombine and remain neutral, especially near the edges of the region where the ionizing background is small.  The interaction of sources and the IGM likely regulates the later stages of reionization and may be important for understanding the properties of the ionized bubbles \cite{furl05-rec, sobacchi14}.

Currently, the most important unknown is the overall ionizing efficiency of the sources responsible for reionization.  Stellar sources are regarded as the most likely candidates. Recent surveys in the Hubble Ultradeep Field suggest that galaxies can, under plausible assumptions, keep the Universe ionized at $z \sim 7$ \cite{robertson13}, but those assumptions require that most of the photons come from galaxies too faint to have been detected to date.  Models generally construct the overall ionizing efficiency with $f_\star f_{\rm esc} N_{\rm ion/b}$, where $f_\star$ is the overall star formation efficiency, $f_{\rm esc}$ is the fraction ionizing photons able to escape their source, and $N_{\rm ion/b}$ is the number of ionizing photons produced per baryon in stars.  Plausible values of $f_\star$ are $< 10\%$. Measurements of $f_{\rm esc}$ in galaxies at $z <4$ also show that $f_{\rm esc} < 10\%$, but there is an expectation in much of the community that the values increase in the smaller galaxies prevalent at $z >6$ \cite{haardt12}.  $N_{\rm ion/b}$ depends on the stellar IMF, metallicity, and other factors like the fraction of binaries, so it is uncertain by a factor of order unity (see the discussion in, e.g., \cite{loeb13}).

There is, as yet, no direct evidence that stars are primarily responsible for reionization, but the only other astrophysically-motivated source -- quasars -- decline very rapidly past $z \sim 4$ \cite{willott10}.  If another population of (faint) accreting black holes existed at high redshifts, they could have profound effects on the ionizing background and the morphology of reionization \cite{ricotti05, volonteri09}.

\begin{figure}[t]
\sidecaption[t]
\includegraphics[width=0.6\textwidth]{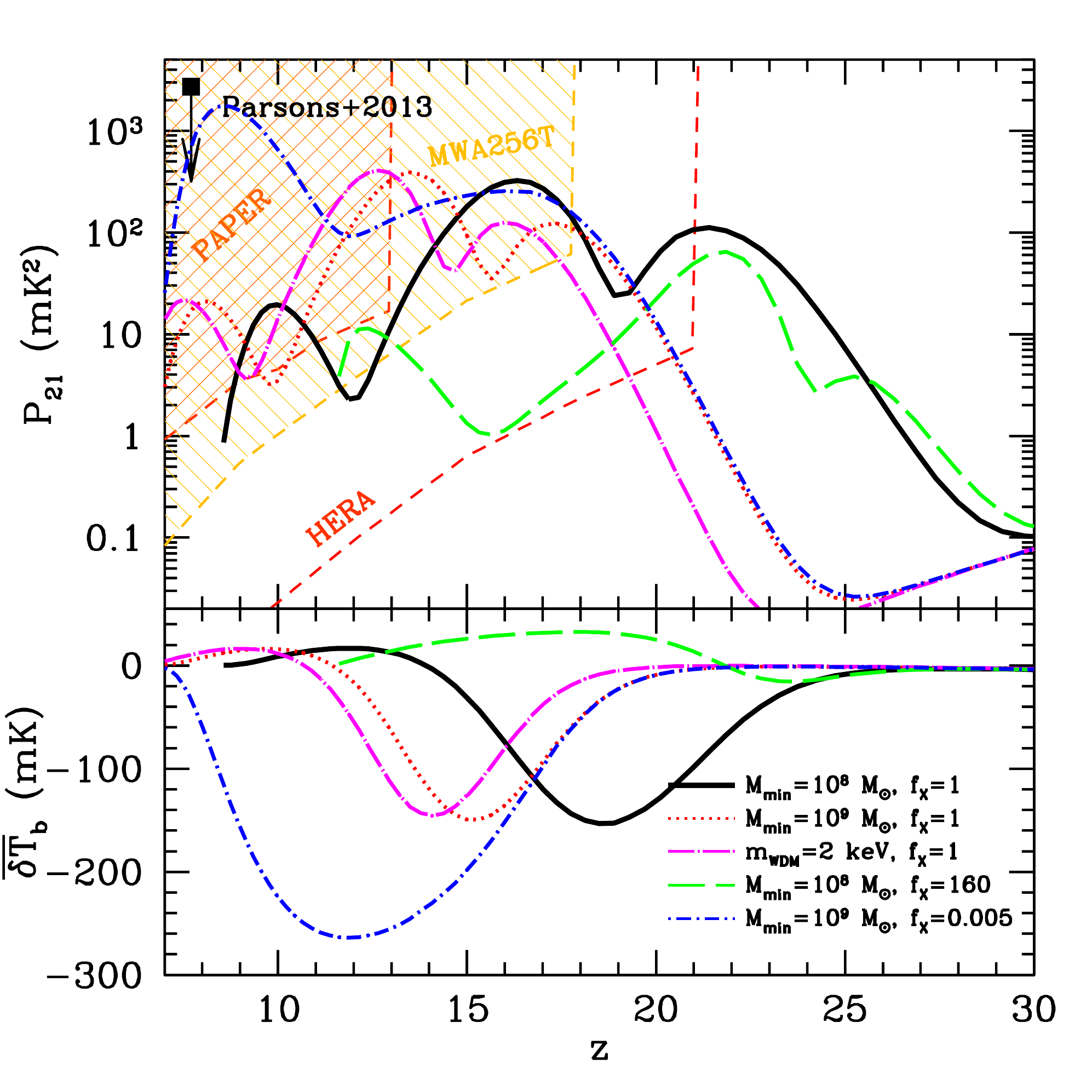}
\caption{{\it Top panel:} Amplitude of the 21-cm power spectrum at $k = 0.1$ Mpc$^{-1}$ in several representative  models (as labeled in the bottom panel).  We also plot the (1$\sigma$) sensitivities of 2000h observation with an expanded MWA with twice the current collecting area, PAPER, and the full HERA array  The recent upper limit from \cite{parsons14} is shown at $z=7.7$. {\it Bottom panel:} The corresponding sky-averaged 21-cm brightness temperature (relative to the CMB) for these models. The input parameters are labeled in the bottom panel.  $M_{\rm min}$ is the minimum halo mass allowed to form stars, $f_X$ is the X-ray efficiency, and the magenta dot-dashed curve uses a cosmology with warm dark matter (which delays structure formation). From \cite{mesinger14}.  
}
\label{fig:sphere-sens}
\end{figure}

\section{The Average Brightness Temperature} \label{tb-monopole}

With the basics in place, we can now compute the time evolution of the brightness temperature $T_b$ in some simple models.  We will begin in this section with the monopole, or sky-averaged brightness, as a function of frequency. The bottom panel of Figure~\ref{fig:sphere-sens} shows the results (as a function of redshift) for several models of early star formation (the upper panel shows the corresponding fluctuations, which we will discuss next).  The principal parameters varied here are $M_{\rm min}$, the minimum halo mass to host star formation, and $f_X$, the X-ray heating efficiency.  We will take the solid black curve as our fiducial model: these choices are simplistic but representative of many models.  Nevertheless, the Figure clearly shows that the signal can plausibly range by orders of magnitude over most of this range.  It also illustrates several important points about the 21-cm background.  The most crucial is the presence of five critical points in the spin-flip background, at least in simple models like this one \cite{furl06-glob,pritchard10-glob}.

\begin{enumerate}

\item The first, at $z \sim 80$, occurs long before star formation becomes significant (and is not shown in this panel).  Over this time, collisional coupling becomes increasingly ineffective, and the turning point occurs roughly when the $\delta_{\rm coll}$ falls below unity (see equation~\ref{eq:dcoll}), at which point $T_S \rightarrow T_{\gamma}$ and the IGM signal begins to fade.  This transition is well-specified by atomic physics and the standard cosmology, at least in the absence of any exotic dark sector processes that may input energy into the IGM at $z \ga 50$.  This signal therefore provides a clear probe of cosmology, at least in principle, but it will be extraordinarily difficult to detect.

\item The remaining transition points are determined by the properties of luminous sources, so their timing is much more uncertain.  In typical models, the next crucial event is the formation of the first stars (at $z \sim 25$), which generate Lyman-$\alpha$ photons and so re-activate the 21-cm background. Interestingly, the timing of this transition is relatively insensitive to the luminosity of these sources, because (at least in this model) the abundance of the massive halos hosting them is increasing so rapidly that their formation is mostly determined by the rate of halo collapse \cite{pritchard10-glob}. 

\item In most models, the next feature is the minimum in $T_b$, which occurs just before IGM heating begins to become significant and is determined primarily by the relative amplitudes of $f_X$, $N_\alpha$, and the ionization efficiency. (If the first is very large, this heating transition can precede strong coupling, while if it is very small it may not occur until reionization is already underway.)  In simple models like we use here, in which both the X-ray and UV luminosities trace $f_{\rm coll}$, the net X-ray heat input $\Delta T_c$ when $x_\alpha=1$ is 
\begin{equation}
\frac{\Delta T_c}{T_\gamma} \sim 0.08 f_X \left( \frac{f_{X,h}}{0.2} \,
\frac{f_{\rm coll}}{\Delta f_{\rm coll}} \, \frac{9690}{N_\alpha} \,
\frac{1}{S_\alpha} \right) \left(
  \frac{20}{1+z} \right)^3,
\label{eq:dtc}
\end{equation}
where $\Delta f_{\rm coll} \sim f_{\rm coll}$ is the effective collapse fraction appearing in the integrals of equation (\ref{eq:jalpha}) and $f_{X,h}$ is the fraction of the X-ray energy that goes into heating (typically $\sim 1/3$; \cite{furl10-xray}).  Note that $\Delta T_c$ is independent of $f_\star$ because we have assumed that both the coupling and heating rates are proportional to the star formation rate.  Clearly, for our fiducial (Population II) parameters the onset of Wouthuysen-Field coupling precedes the point at which heating begins, which is ultimately the reason for the strong absorption in our fiducial model. 

\item The fourth turning point occurs at the maximum of $T_b$.  In the fiducial model, this marks the point at which $T_K \gg T_{\gamma}$, so that the temperature portion of equation~(\ref{eq:dtb}) saturates.  The signal then starts to decrease rapidly once reionization begins in earnest.  Most likely, this happens \emph{after} coupling is already strong and heating is significant. Again, in the simple models used here the ionized
fraction at $x_\alpha=1$ is given by
\begin{equation}
\bar{x}_{i,c} \sim 0.05 \left( \frac{f_{\rm esc}}{1+\bar{n}_{\rm rec}} \,
\frac{N_{\rm ion}}{N_\alpha} \, \frac{f_{\rm coll}}{\Delta f_{\rm coll}} \,
\frac{1}{S_\alpha} \, \right) \left( \frac{20}{1+z} \right)^2,
\label{eq:xic}
\end{equation}
where $\bar{n}_{\rm rec}$ is the mean number of recombinations per baryon. For Population II stars with a normal IMF, $N_{\rm ion}/N_\alpha \approx 0.4$ \cite{barkana05-ts}; thus, even in the worst case of $f_{\rm esc}=1$ and $\bar{n}_{\rm rec}=0$ coupling would become efficient during the initial stages of reionization.  However, very massive Population III stars have much harder spectra, with $N_{\rm ion}/N_\alpha \approx 7$. In principle, it is therefore possible for Pop III stars to reionize the universe \emph{before} $x_\alpha=1$, although this is rather unlikely given their fragility.

Whether the IGM will appear in absorption or emission during reionization is more controversial.  We find
\begin{equation}
\frac{\Delta T}{T_\gamma} \sim \left( \frac{\bar{x}_i}{0.025} \right) \,
\left( f_X \, \frac{f_{X,h}}{f_{\rm esc}} \, \frac{4800}{N_{\rm ion}} \,
\frac{10}{1+z} \right) \, (1 + \bar{n}_{\rm rec})
\label{eq:tx-xi}
\end{equation}
for the heat input $\Delta T$ as a function of $\bar{x}_i$. Thus, provided $f_X \ga 1$, the IGM will be much
warmer than the CMB during the bulk of reionization.  But this is by no means assured, and some models predict that the IGM will remain cold until the midpoint of reionization \cite{fialkov14}.

\item The monopole signal (nearly) vanishes when reionization completes.

\end{enumerate}

Several efforts to observe this monopole signal are either completed or now underway, including the Cosmological Reionization Experiment (CoRE), the Experiment to Detect the Global Epoch of Reionization Signal (EDGES)\footnote{See http://www.haystack.mit.edu/ast/arrays/Edges/} \cite{bowman10}, the SCI-HI experiment \cite{voytek14}, the Large Aperture Experiment to Detect the Dark Ages (LEDA)\footnote{http://www.cfa.harvard.edu/LEDA/}, and an ambitious program to launch a radio telescope to the moon in order to observe the high-redshift signal is also being planned (the Dark Ages Radio Telescope, or DARE)\footnote{http://lunar.colorado.edu/dare/} \cite{burns12}.  

Because global experiments aim to detect an all-sky signal,
single-dish measurements (even with a modest-sized telescope) can
easily reach the required mK sensitivity \cite{shaver99}.  However, the much stronger
synchrotron foregrounds from our Galaxy nevertheless make such
observations extremely difficult: they have $T_{\rm sky} \ga
200$--$10^4 \kel$ over the relevant frequencies (see the map in
Figure~\ref{fig:21cm-foreground}).  The fundamental strategy for
extracting the cosmological signal relies on the expected spectral
smoothness of the foregrounds (which primarily have power law
synchrotron spectra), in contrast to the non-trivial structure of the
21-cm background.  Nevertheless, isolating the high-redshift
component will be a challenge that requires extremely accurate
calibration over a wide frequency range and, most likely, sharp
localized features in ${T}_b(z)$ that can be distinguished from
smoother foreground features.

Current estimates show that rapid reionization histories which span a redshift range $\Delta z\lesssim 2$ can be constrained, provided that local foregrounds can be well modeled \cite{bowman10}.  Observations in the frequency range 50-100~MHz can potentially constrain the Lyman-$\alpha$ and X-ray emissivity of the first stars and black holes: even though the foregrounds are significantly worse at these lower frequencies, the strong absorption signal present in many models may be easier to observe than the gently-varying reionization signal.  However, it may be necessary to perform such observations from space, in order to avoid systematics from terrestrial interference and the ionosphere, whose properties strongly vary spatially, temporally, and with frequency (in particular, the ionosphere crosses from absorption to emission in this range; \cite{datta14}). In fact the best observing environment is the far side of the moon (though also the most expensive!), where the moon itself blocks any radio signals from Earth; this is the primary motivation for DARE.

\begin{figure}[!t]
\centerline{\includegraphics[scale=0.3,angle=-90]{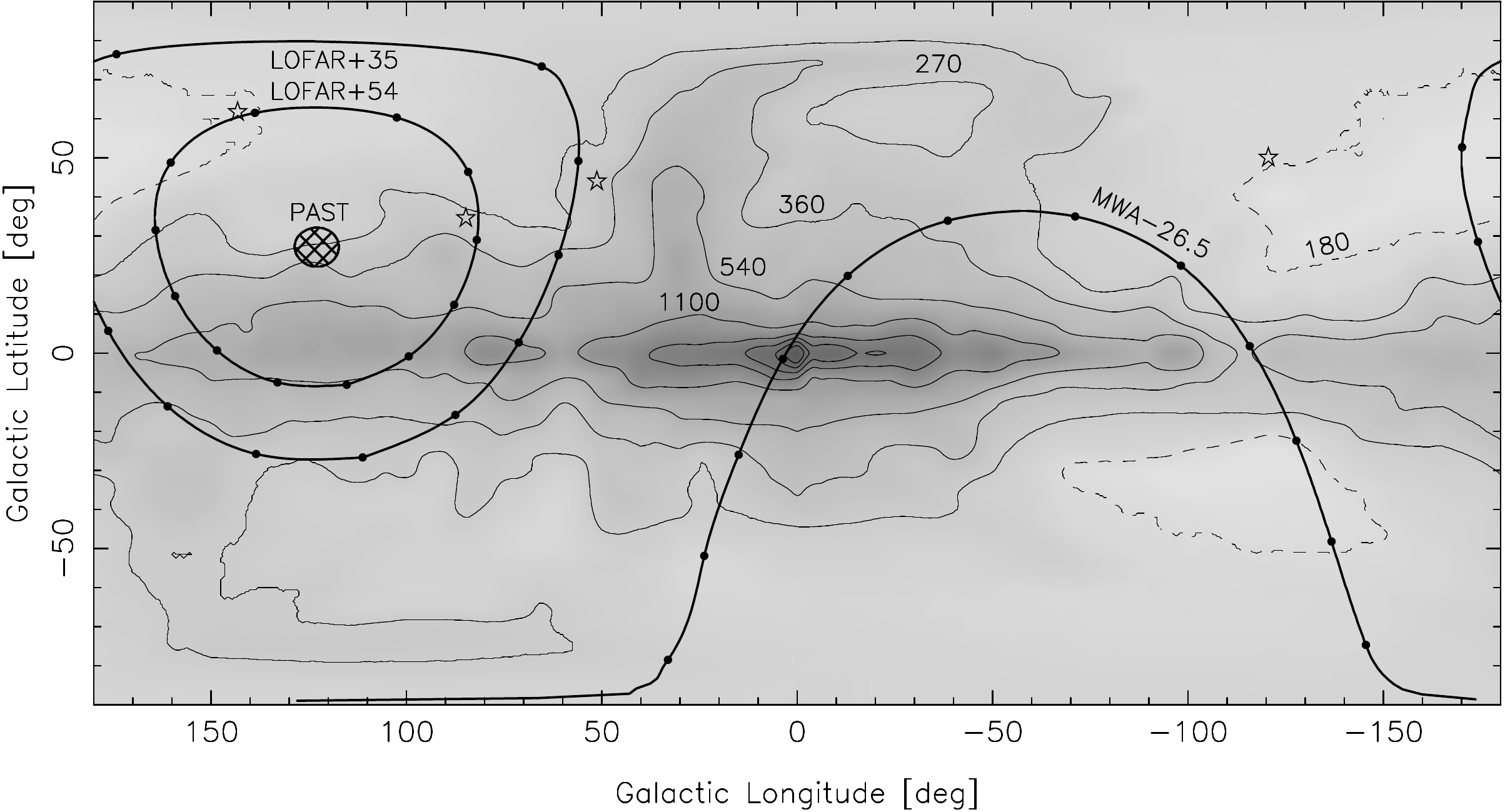}}
\caption{Brightness temperature of the radio sky at 150 MHz
in Galactic coordinates.  Contours are drawn at 180
(dashed), 270, 360, 540, 1100, 2200, 3300, 4400, and 5500 K.  A potential survey field at the North celestial pole is
cross-hatched. Heavy lines indicate constant
declinations:$-26.5^{\circ}$, $+35^{\circ}$, and $+54^{\circ}$ with
dots to mark 2 hour intervals of time (these are ideal for two other
existing experiments, the Murchison Wide-field Array or MWA and
LOFAR). Star symbols indicate the coordinates of four bright $z>6.2$
quasars. From \cite{furl06-review}, based on data in \cite{landecker69}. Copyright 2006 by Elsevier.}
\label{fig:21cm-foreground}
\end{figure}

\section{Statistical Fluctuations in the Spin-Flip Background} \label{21-flucs}

While the 21~cm monopole contains a great deal of information about the mean evolution of the sources, every component in equation~(\ref{eq:dtb}) can also fluctuate significantly.  The evolving cosmic web imprints growing density fluctuations on the matter distribution.  Ionized gas is organized into discrete H II regions (at least in the most plausible models), and the \lya background and X-ray heating will also be concentrated around galaxies.  The single greatest advantage of the 21-cm line is that it allows us to separate this fluctuating component both on the sky and in frequency (and hence cosmic time).  Thus, we can study the sources and their effects on the IGM in detail.  It is the promise of these ``tomographic" observations that makes the 21 cm line such a singularly attractive probe.

Observing the 21-cm fluctuations has one practical advantage as well. The difficulty of extracting the global evolution from the enormously bright foregrounds shown in Figure~\ref{fig:21cm-foreground} lies in its relatively slow variation with frequency.  On the small scales relevant to fluctuations in the signal, the gradients increase dramatically: for example, at the edge of an H II region $T_b$ drops by $\sim 20 \mkel$ essentially instantaneously.  As a result, separating them from the smoothly varying astronomical foregrounds may be much easier.  Unfortunately, constructing detailed images will remain extremely difficult because of their extraordinary faintness; telescope noise is comparable to or exceeds the signal except on rather large scales (see \S \ref{21cm-obs} below).  Thus, a great deal of attention has recently focused on using statistical quantities readily extractable from low signal-to-noise maps to constrain the IGM properties.  This is motivated in part by the success of CMB measurements and galaxy surveys at constraining cosmological parameters through the power spectrum.\footnote{Other statistical measures, such as higher-order correlations, may also offer additional information.}  

We first define the fractional perturbation to the brightness temperature, $\delta_{21}({\bf x}) \equiv [T_b({\bf x}) - \bar{T}_b]/\bar{T}_b$, a zero-mean random field.  We will be interested in its Fourier transform $\tilde{\delta}_{21}({\bk})$. Its power spectrum is defined to be
\begin{equation}
\VEV{ \tilde{\delta}_{21}(\bk_1) \, \tilde{\delta}_{21}(\bk_2) } \equiv (2 \pi)^3 \delta_D(\bk_1 - \bk_2) P_{21}(\bk_1),
\label{eq:pkdefn}
\end{equation}
where $\delta_D(x)$ is the Dirac delta function and the angular brackets denote an ensemble average.  Power spectra for other random fields (such as the fractional overdensity $\delta$, the ionized fraction, etc.), or cross-power spectra between two different fields, can be defined in an analogous fashion.

Expanding equations~(\ref{eq:dtb}) and (\ref{eq:xdefn}) to linear order in each of the perturbations, we can write 
\begin{equation}
\delta_{21} = \beta \delta_b + \beta_x \delta_x + \beta_\alpha \delta_\alpha + \beta_T \delta_T -  \delta_{\partial v},
\label{eq:d21}
\end{equation}
where each $\delta_i$ describes the fractional variation in a particular quantity: $\delta_b$ for the baryonic density (for which the total density is an adequate approximation on large scales), $\delta_\alpha$ for the \lya coupling coefficient $x_\alpha$, $\delta_x$ for the neutral fraction, $\delta_T$ for $T_K$, and  $\delta_{\partial v}$ for the line-of-sight peculiar velocity gradient.  The expansion coefficients $\beta_i$ can be written explicitly \cite{barkana05-ts}; for example,
\begin{eqnarray}
\beta & = & 1 + \frac{x_c}{x_{\rm tot}(1+x_{\rm tot})},
\label{eq:beta} \\
\beta_\alpha & = & \frac{x_\alpha}{x_{\rm tot}(1+x_{\rm tot})},
\label{eq:beta-alpha} \\
\end{eqnarray}
where $x_{\rm tot} \equiv x_c + x_\alpha$. These expressions have simple physical interpretations.  For $\beta$, the first term describes the increased matter content and the second describes the increased collisional coupling efficiency in dense gas, while $\beta_\alpha$ simply measures the fractional contribution of the Wouthuysen-Field effect to the coupling.  By linearity, the Fourier transform $\tilde{\delta}_{21}$ can be written in a similar fashion.   

Based on equation~(\ref{eq:pkdefn}), the power spectrum contains all
possible terms of the form $P_{\delta_i \delta_j}$; some or all could
be relevant in any given situation.  Of course, in most circumstances the
various $\delta_i$ will be correlated in some way; statistical 21 cm
observations ideally hope to measure these separate quantities.  

In all of these expansions, one must bear in mind that $\delta_x$ is \emph{always} of order unity if the ionization field is built from H~II regions, because $x_i=0$ or $1$.  In that case terms such as $\delta \delta_x$ are in fact \emph{first} order and must be retained in detailed calculations \cite{mcquinn06-param}.  This is a general limitation of the linear theory approach in equation~(\ref{eq:d21}), as nonlinear effects play a very important role during reionization and sometimes before

In general, we expect the fluctuations in density, ionization fraction, Ly$\alpha$ flux, and temperature to be statistically isotropic, because the physical processes responsible for them have no preferred direction [e.g., $\delta(\bk) = \delta(k)$].  However, peculiar velocity gradients introduce anisotropic distortions.  Bulk flows on large scales, and in particular infall onto massive structures, compress the signal in redshift space (the so-called \emph{Kaiser effect}; \cite{kaiser84}),  enhancing the apparent clustering amplitude. On small scales, random motions in virialized regions create elongation in redshift space (the ``finger of God" effect), reducing the apparent clustering amplitude (though only on scales irrelevant to 21-cm observations). If we label the coordinates in redshift space by ${\bf s}$, it is straightforward to show that \cite{kaiser84}
\beq 
\delta_s({\bf k}) =
\delta({\bf k})[ 1 + \beta \mu_{\bf k}^2] 
\eeq 
where $\mu_{\bf k} = \hat{\bf k} \cdot \hat{\bf x}$ is the cosine of the angle between the
wave vector and the line of sight, $\beta \approx \Omega_m^{0.6}(z)$ corrects for a possible bias between the
tracers we are studying and the growth rate of dark matter perturbations. 

The redshift-space distortions therefore provide an anisotropic \emph{amplification} to the background signal \cite{bharadwaj04-vel}.  The anisotropy occurs because only modes along the line of sight are affected.  To understand the amplification, consider a spherical overdense region. Its excess gravitational force causes it to recollapse.  Along the radial direction, the collapse \emph{decreases} the velocity width of the object relative to the Hubble flow (at least in linear theory), compressing the overdensity in redshift space.  Similarly, a spherical underdensity expands faster than average, causing it to appear elongated in the radial direction.  Averaged over all modes, these distortions amplify the signal by a factor $\approx \VEV{(1+ \mu^2)^2} \approx 1.87$.

However, the anisotropies are actually even more helpful in that they provide angular structure to the signal, which may allow us to separate the many contributions to the total power spectrum. Schematically, brightness temperature fluctuations in Fourier space have the form \cite{barkana05-vel}
\begin{equation}
\delta_{21} =  \mu^{2} \beta \delta + \delta_{\rm iso} 
\label{eq:d21ft}
\end{equation}
where we have collected all the statistically isotropic terms in
equation~(\ref{eq:d21}) into $\delta_{\rm iso}$. Neglecting
``second-order" terms (see below) and setting $\beta=1$, the total
power spectrum can therefore be written as \cite{barkana05-vel} 
\begin{equation}
P_{21}({\bf k}) = \mu^{4} P_{{\delta} {\delta}} + 2 \mu^{2} P_{{\delta}_{\rm iso} {\delta}} + P_{{\delta}_{\rm iso} {\delta}_{\rm iso}}.
\label{eqn:p_polynomial}
\end{equation}
By separately measuring these three angular components (which requires, in principle, estimates at just a few values of $\mu$), we can in principle isolate the contribution from density fluctuations $P_{\delta \delta}$.  This would not have been possible without peculiar velocity flows:  comparison to equation~(\ref{eq:d21}) shows that, in the most general case, $P_{{\delta}_{\rm iso} {\delta}}$ and $P_{{\delta}_{\rm iso} {\delta}_{\rm iso}}$ contain several different power spectra, including those of the density, neutral fraction, and spin temperature as well as their cross power spectra. However, in practice nonlinear evolution and/or the combination of other effects can dominate the behavior quite easily \cite{mesinger11, mao12, ghara14}. It is not yet clear how useful these redshift space distortions will be in practice.

\section{Spin-Flip Fluctuations During the Cosmic Dawn} \label{21-flucs-history}

Figure~\ref{fig:21cmslices} shows several snapshots of a ``semi-numerical" computer simulation of the spin-flip background (essentially, a realization of a model universe using linear theory to determine the locations of luminous sources and ionized bubbles and the framework described above to calculate the inhomogeneous radiation backgrounds; \cite{mesinger07, mesinger11}), including both snapshots of the fields (in the left column) and the corresponding (spherically-averaged) power spectra (in the right column).  The underlying model is very similar to the fiducial model whose mean signal is shown in Figure~\ref{fig:sphere-sens}, though the redshifts of the critical points differ slightly.  

\begin{figure}[!t]
\centerline{\includegraphics[width=0.6\textwidth]{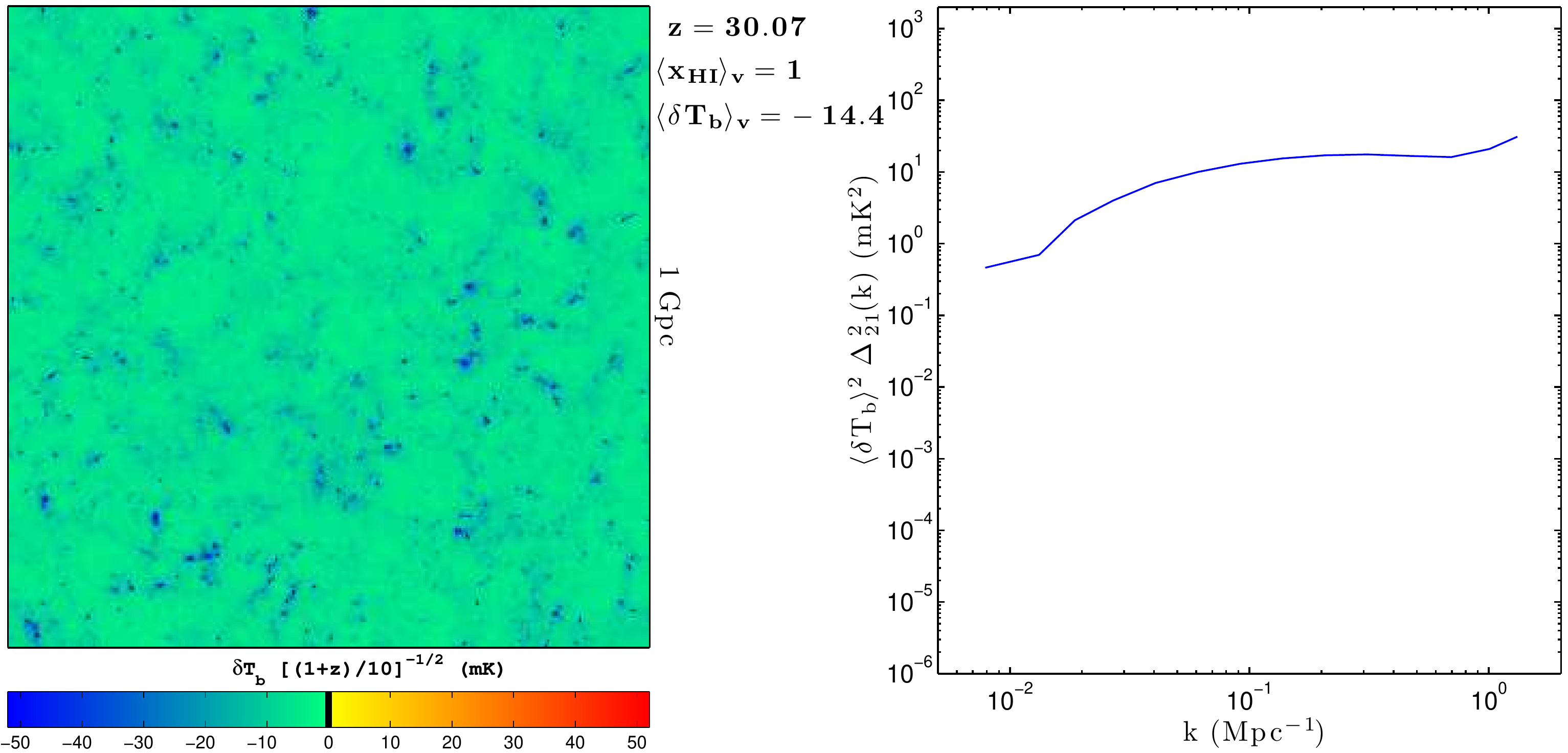}}
\centerline{\includegraphics[width=0.6\textwidth]{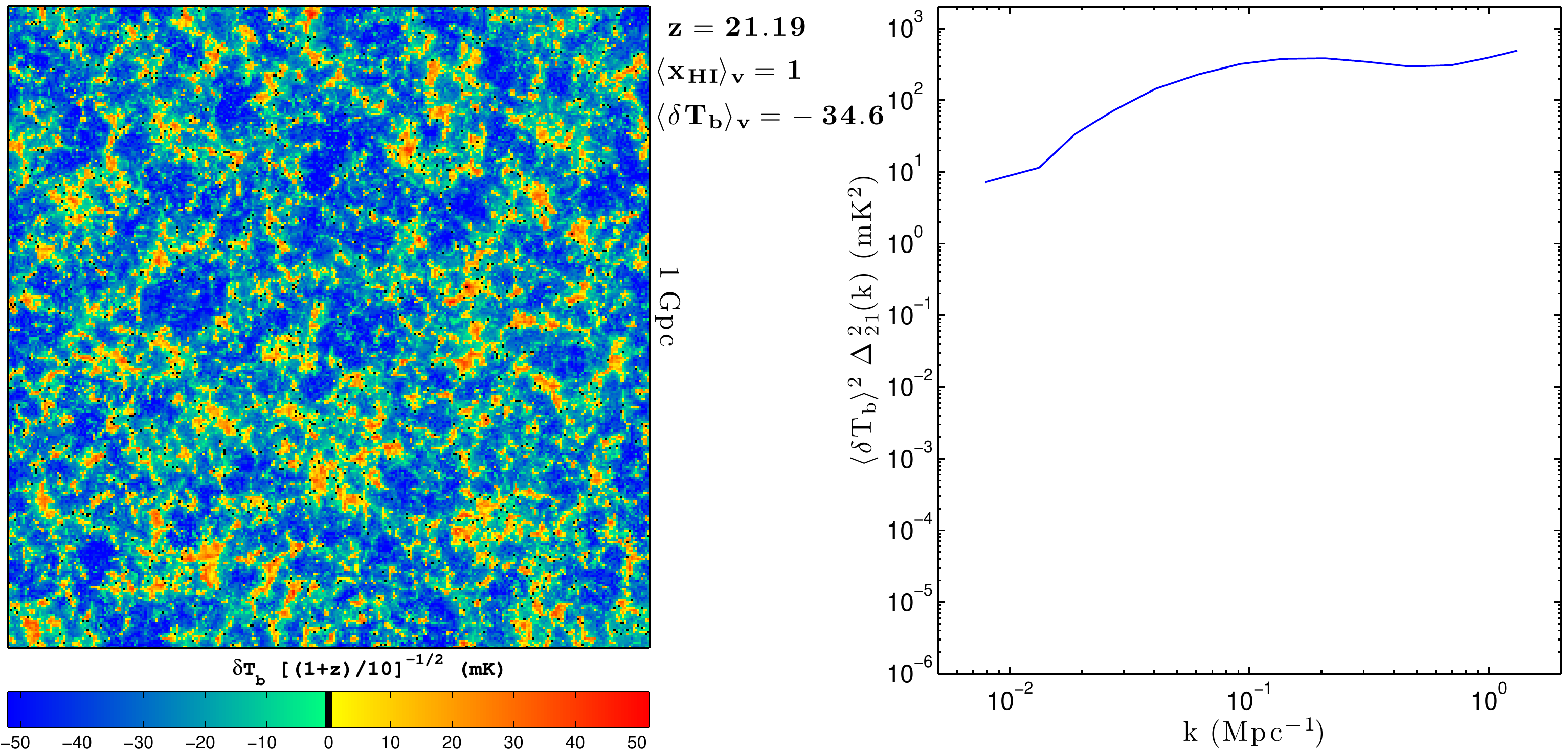}}
\centerline{\includegraphics[width=0.6\textwidth]{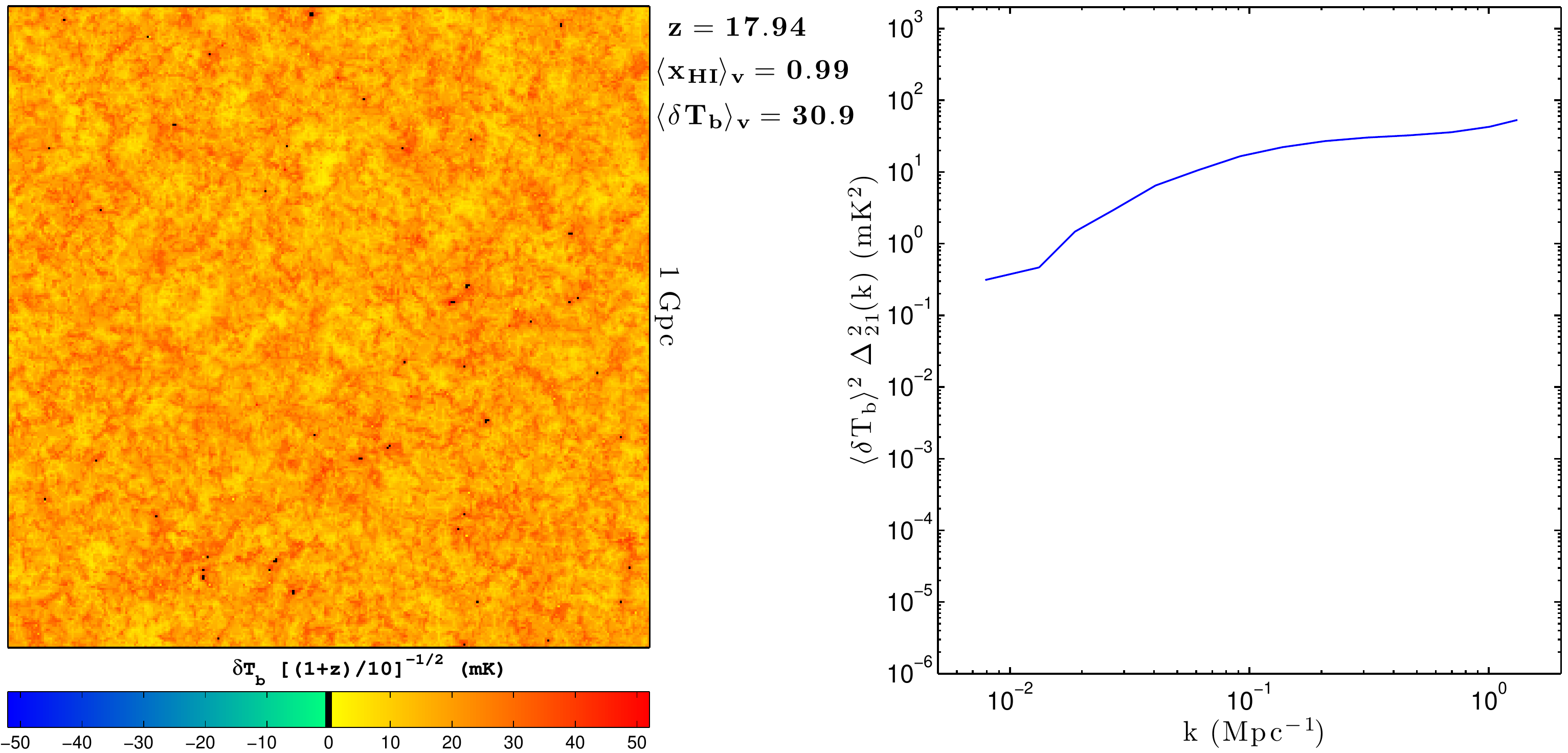}}
\centerline{\includegraphics[width=0.6\textwidth]{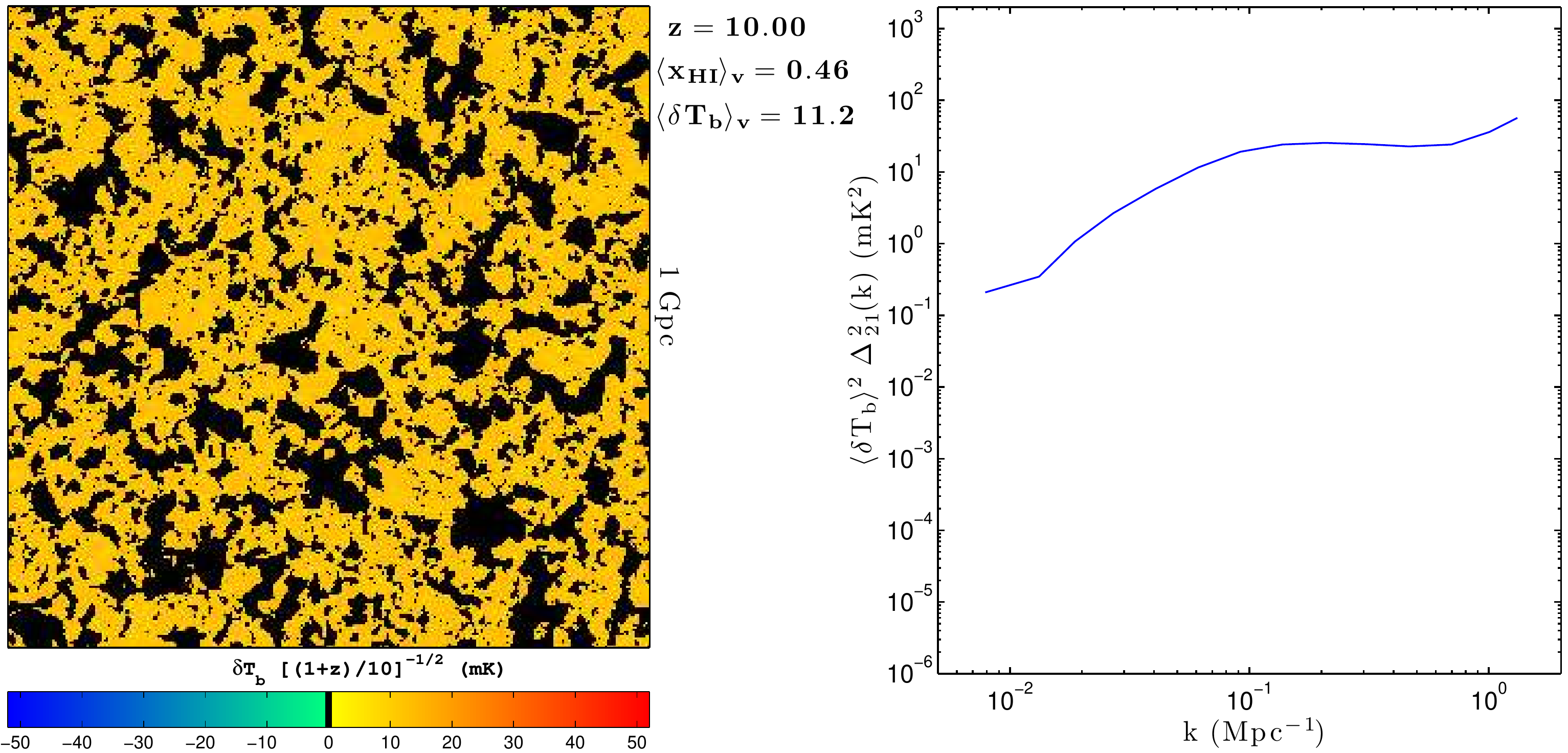}}
\caption{Slices through a ``semi-numerical" simulation ({\it left}), and
the corresponding spherically-averaged power spectra ({\it right}),
for a model of the spin-flip background at $z = $ 30.1, 21.2, 17.9,
10.0 ({\it top to bottom}).  The slices were chosen to highlight
various epochs in the cosmic 21-cm signal (from top to bottom): the onset of Lyman-$\alpha$
pumping (here the blue regions show the cold gas around the first
galaxies), the onset of X-ray heating (here the blue regions
are  cold gas, while the compact red regions
represent hot gas around the first black holes), the completion of
X-ray heating (where all the gas is hot), and the mid-point of
reionization (where black regions are ionized bubbles).  All comoving slices are 1 Gpc on a side and 3.3 Mpc
deep.  From \cite{mesinger11}. Copyright 2011 by the Royal Astronomical Society.}
\label{fig:21cmslices}
\end{figure}

The top row of Figure~\ref{fig:21cmslices} shows the point where Lyman-$\alpha$ pumping begins to be significant.  The hydrogen gas is cold ($T_K \ll T_{\gamma}$), and the spin temperature is just beginning to decouple from the CMB.  In this case the fluctuations are driven by the discrete, clustered first galaxies: their radiation field drives $T_S \rightarrow T_K$ around those first sources, while leaving most of the IGM transparent.

In this calculation, the Lyman-$\alpha$ radiation field very quickly builds up the brightness temperature fluctuations. We also illustrate this in the top panel of Figure~\ref{fig:sphere-sens}, which shows the evolution of the amplitude of the power spectrum at one particular wavenumber ($k=0.1 \Mpcinv$, near the peak sensitivities of most arrays). The rightmost peak of the solid curve shows the effects of the Lyman-$\alpha$ fluctuations: they build up to a peak, with amplitude $\sim 10$~mK, before decreasing again once the Lyman-$\alpha$ background becomes strong throughout the universe (at which point $\beta_\alpha \propto 1/x_\alpha \rightarrow 0$).

The second row in Figure~\ref{fig:21cmslices} shows a map shortly after X-ray heating commences. At this point in the model, the Lyman-$\alpha$ coupling is strong nearly everywhere, so most of the IGM appears in absorption.  But near the first X-ray sources, the gas has $T_S \gg T_{\gamma}$, so these regions appear in
emission.  The net effect is a very large fluctuation amplitude, with a strong contrast between emitting and absorbing regions, as we see in the middle peak of the solid curve in Figure~\ref{fig:sphere-sens}.  

The third row in Figure~\ref{fig:21cmslices} shows the 21-cm signal after heating has saturated ($T_S \gg T_{\gamma}$) throughout the IGM. At this point, spin temperature fluctuations no longer contribute to $T_b$, and only the density field affects the overall signal.  The fluctuations are thus relatively modest (also seen in Figure~\ref{fig:sphere-sens}).  (Note that this is a feature of our parameter choices here: if X-ray heating is weaker, it can overlap with reionization, mixing the types of fluctuations.)

Finally, the fluctuations increase again once reionization begins in earnest, as shown in the bottom row of Figure~\ref{fig:21cmslices}: here the contrast between the ionized bubbles and fully neutral gas in between them dominates the features.  These bubbles are the key observable during reionization, as their pattern depends on the properties of the ionizing sources.

The other curves in the top panel of Figure~\ref{fig:sphere-sens} show how the fluctuations on this scale can vary in a plausible range of models.  Note that most provide the same overall structure,with three consecutive peaks, but their timing and amplitudes vary.  Moreover, the blue short dashed-dotted curve, which assumes very weak X-ray heating, has only two peaks, as the IGM is not heated substantially until reionization is already underway.  The broad range of possible signals makes the 21-cm line a powerful probe.

Figure~\ref{fig:reion-ps} shows the evolution of the power spectrum during reionization in considerably more detail (it assumes $T_S \gg T_\gamma$ throughout) \cite{lidz08-constraint}.  In particular, it plots the dimensionless power spectrum $\Delta^2_{21}(k) = k^3 P_{21}(k)/(2 \pi^2)$ (or the power per logarithmic interval in wavenumber of the 21-cm signal) over the course of a radiative transfer simulation of the reionization process. (To recover the 21-cm signal one needs to multiply these values by the mean brightness temperature in a fully neutral medium, $T_0^2 \approx [28^2 (1+z)/10]$~mK$^2$.) The different curves show a sequence of ionized fractions, from nearly neutral ($\VEV{x_i}=0.02$) to almost fully ionized ($\VEV{x_i}=0.96$).  In this model, these go from $z \sim 11.5$--$6.8$, but the curves change little if one holds $\VEV{x_i}$ constant but chooses a different redshift.

\begin{figure}[!t]
\sidecaption[t]
\includegraphics[width=0.6\textwidth]{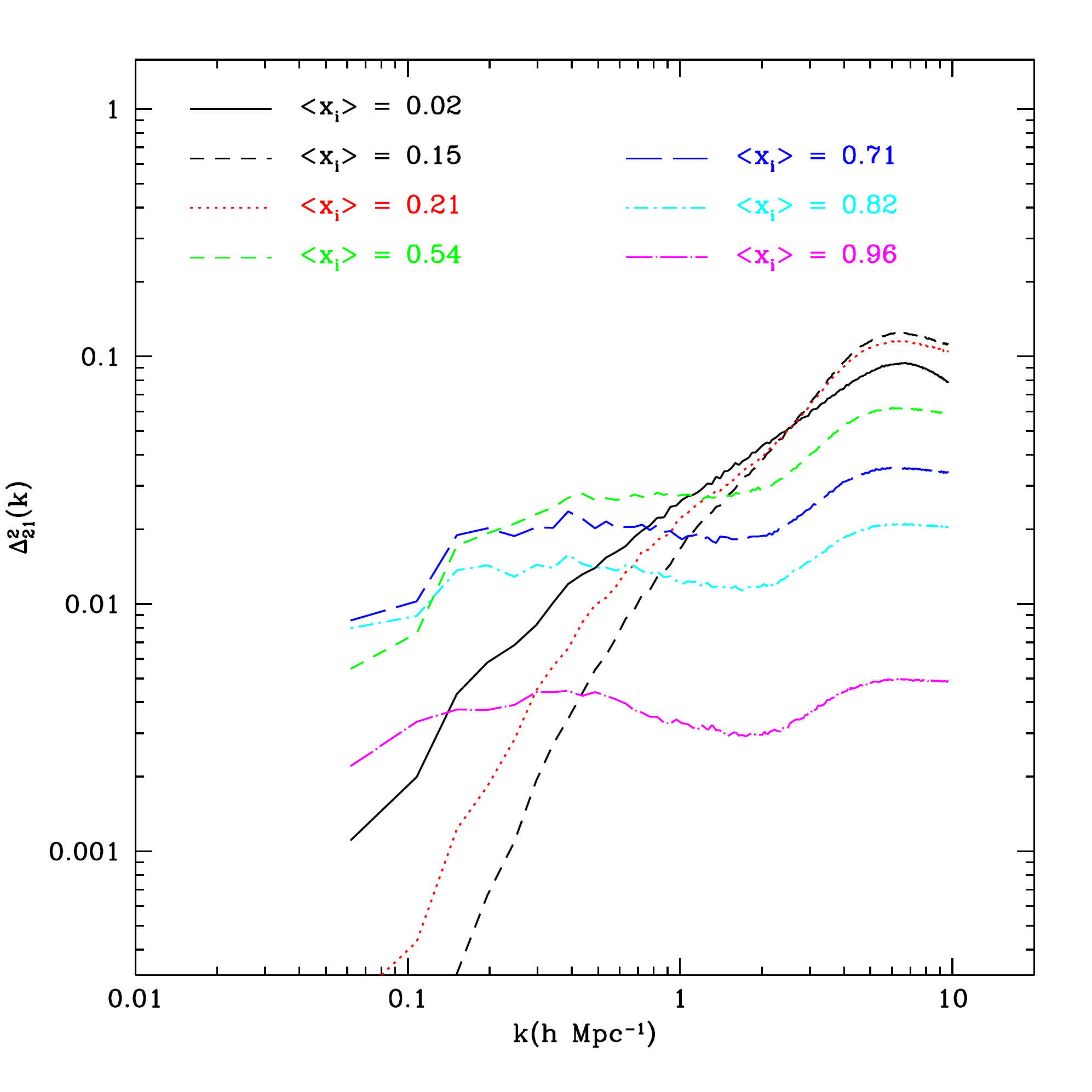}
\caption{Dimensionless power spectra $\Delta^2_{21}(k)$ of spin-flip
background during the reionization era in a numerical simulation. The curves show the power
spectrum through a sequence of mean ionized fractions; the redshifts
at which these points are achieved (not listed) do not significantly
affect the signal, except through the mean brightness temperature. From \cite{lidz08-constraint}. Reproduced with permission of the American Astronomical Society.}
\label{fig:reion-ps}
\end{figure}

At first, the 21-cm power spectrum simply traces the matter power spectrum, as ionized regions have not yet significantly affected the IGM. But fairly quickly, the power \emph{decreases} on large scales because the ionized bubbles appear first in the densest regions, suppressing the signal there and hence decreasing the overall contrast in the 21-cm maps.

This is simplest to understand if we decompose the power spectrum into
parts that describe perturbations in each relevant physical parameter
and retain only the terms that dominate during reionization (see equation~\ref{eq:d21}) \cite{lidz07-ng}
\beq
\Delta_{21}^2(k) = T_0^2 \VEV{x_H}^2 \left[ \Delta^2_{\delta
\delta}(k) + 2 \Delta^2_{x \delta}(k) + \Delta^2_{x x}(k) \right].
\label{eq:d21-decomp}
\eeq 
Here, $\Delta^2_{\delta \delta}$ and $\Delta^2_{x x}$ are the dimensionless power spectra of the density field and ionized fraction, and $\Delta^2_{x \delta}$ is the cross-power spectrum of these two quantities. Because $\Delta^2_{x \delta}$ is a cross-power, it can be negative -- i.e., the neutral fraction $x_H$ is small when $\delta$ is large in most reionization models.  In the early phases of reionization, this term dominates the ionized power itself, $\Delta^2_{xx}$, and so the net power declines as $x_H$ initially increases.

However, by $\VEV{x_i} \sim 0.5$, the $\sim 20$~mK contrast between ionized and neutral gas dominates the maps, and the power increases rapidly on large scales: now the ionized bubbles fill a wide range of density, so $\Delta^2_{xx} \gg \Delta^2_{x \delta}$.  The power from this term peaks on the characteristic scale of the ionized bubbles. In combination with the contribution from the matter power spectrum itself, this leads to a strong enhancement of power on moderate scales ($k \sim 0.1 \Mpcinv$), followed by a decline at smaller wavenumbers (not shown clearly in this figure because of the finite size of the simulation box).

At the same time, on scales much smaller than the bubble size, the
21-cm power is significantly smaller than expected from the matter
power spectrum alone.  This is largely because of the higher-order
terms that we have ignored: within an ionized region, the ionized
fraction is uncorrelated with the small-scale density
perturbations.  Effectively then the contrast on these scales is
decreased because many of the small-scale overdensities no longer
appear in the 21-cm map.  The net effect is an overall
\emph{flattening} in $\Delta_{21}^2$ throughout reionization. The
flattening shifts to larger scales as the process unfolds. Meanwhile, the
overall amplitude of the signal decreases as less of the gas can emit 21-cm photons.

\begin{figure}[t]
\sidecaption[t]
\includegraphics[width=0.6\textwidth]{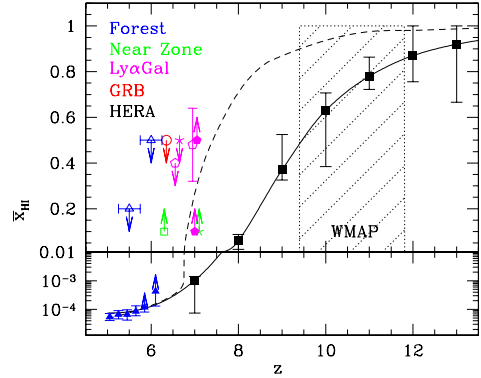}
\caption{Existing constraints on neutral fraction ($x_{\rm HI}$) versus redshift (adapted from \cite{robertson13}) along with a pair of fiducial reionization histories (black lines). The solid curve with error bars is a typical theoretical model of reionization consistent with WMAP measurements, while the dashed line is estimated from galaxy measurements at $z < 8$ and is consistent with most other constraints. The black markers with error bars show predicted HERA-331 constraints on the former model. Courtesy HERA Collaboration.}
\vskip -1.2in
\label{fig:sens-xHI}
\end{figure}

\section{Observing the Spin-Flip Background} \label{21cm-obs}

The potential rewards of studying the early phases of galaxy formation with the 21-cm spectral line -- illustrated as predicted constraints on the reionization history in Figure~\ref{fig:sens-xHI} -- have motivated the construction of several arrays of low-frequency radio antennae over the past several years, with plans for bigger and better instruments in the near future.  For redshifts $z \sim 6$--$50$, the corresponding observed frequencies are $\nu_{\rm obs} \sim 30$--$200$~MHz. The technology for such telescopes has existed for decades -- and is essentially the same that we use every day for TV or radio communication. In fact, efforts to detect the 21-cm background have been made several times over the last half-century (e.g., \cite{field59-obs, davies78, bebb86, uson91a, uson91b}).  However -- in addition to the challenge of having little theoretical guidance -- these early
experiments failed because of three obstacles that still challenge us today: 

\begin{itemize}

\item This band is heavily used by humans (as it includes the FM radio band, analog TV stations, and a host of satellite and aircraft communications channels), and the resulting  radio frequency interference (or RFI) is as many as ten orders of magnitude brighter than the 21-cm background.  Most of the efforts therefore place the
observatories in isolated locations far from the contaminating sources, though some rely on excising the interference from narrow frequency bands. Even then, the presence of such bright foregrounds places
serious requirements on the dynamic range of the low-frequency
observatories.

\item The {\it ionosphere} is refractive over most of this band and opaque at the lower frequencies.  This causes sources to jitter across the sky as patches of
the ionosphere move across the telescope beam.  The refraction
phenomenon is similar to atmospheric seeing in optical astronomy, although the
timescale for the jitter is much slower (several seconds in this
case).  It can be corrected in software by calibrating to the
locations of a set of point sources distributed across the field of
view, although this is by no means a trivial computing effort. The
ionosphere is more active during the day and during times of high
solar activity. This -- together with the large brightness of the sun
itself at these frequencies -- restricts observations to the nighttime hours.

\item Most significantly, the spin-flip background is far from the
only astronomical source in the sky. Nearly all non-thermal radio
sources are bright in the low-frequency band, especially the
synchrotron radiation from the Milky Way galaxy, as we have already
seen in Figure~\ref{fig:21cm-foreground}.  But other extragalactic
sources -- including AGN, galaxy clusters, and even normal
star-forming galaxies -- also contribute.  As a rule of thumb, typical
high-latitude, ``quiet" portions of the sky have a brightness temperature \cite{furl06-review}
\begin{equation}
T_{\rm sky} \approx 180 \ \left( \frac{\nu}{180 \MHz} \right)^{-2.6} \kel.
\label{eq:tsky}
\end{equation}
We immediately see that 21-cm mapping will require large integration times and large collecting area to overcome this ``noise,", which is at least $10^4$ times stronger than the reionization signal.

\end{itemize}

Currently, several experiments are either in the early phases of operations or final phases of construction, including:

\begin{itemize}

\item The Giant Metrewave Radio Telescope (GMRT; in India) is an array of thirty 45-m antennas operating at low radio frequencies.  This large collecting area makes it a powerful instrument, but the instrument's narrow field of view and difficult radio environment present challenges. Nevertheless, the GMRT team was the first to put limits on the spin-flip background in the summer of 2010 \cite{paciga11, paciga13}.

\item The Low Frequency Array (LOFAR; with the core in the Netherlands and outlying stations throughout Europe) is a large, general-purpose low-frequency radio telescope that began science operations in 2010. While its many other science goals mean that LOFAR is not completely optimized to observe the spin-flip background, its large collecting area (especially inside a compact ``core" most useful for these observations) makes it a powerful machine for this purpose. Its location in Western Europe means that LOFAR will face by far the most difficult terrestrial radio environment. Moreover, it uses an enormous number of dipole antennae, combining their individual signals into ``stations" that are then used as interferometers. While this allows for a large collecting area, it presents analysis challenges in understanding the instruments sufficiently well to extract the tiny cosmological signal.

\item The Murchison Widefield Array (MWA) in Western Australia is an
interferometer built almost entirely to observe the 21-cm background.
As such, the project hopes to leverage the relatively small experiment
into limits competitive with larger first-generation experiments.
Like LOFAR, MWA uses thousands of dipoles grouped into ``tiles," which
increase the collecting area at the cost of complexity.  Because MWA's
tiles are smaller, though, it achieves a larger field of view than
LOFAR, which partially compensates for the much smaller collecting area. 

\item The Precision Array to Probe the Epoch of Reionization (PAPER, with instruments in Green Bank, West Virginia and South Africa) combines signals from single dipoles into an interferometer.  Without tiles, PAPER has a much smaller total collecting area than the other efforts but the advantages of a well-calibrated and well-understood instrument and an enormous field of view.  It placed the first physically relevant limits on the IGM at $z \sim 8$ in 2013 \cite{parsons14}, ruling out a cold, neutral IGM at that time (shown in Figure~\ref{fig:sphere-sens}).

\end{itemize}

In addition to this impressive suite of ongoing efforts, larger experiments are planned for the future, with their designs  and strategies informed by this present generation. These include the Hydrogen Epoch of Reionization Array (HERA), which will eventually have hundreds of 14-m dishes optimized to use the strategies developed to analyze PAPER and MWA data (currently beginning the first stage of the array, with an eye toward completion by the end of the decade) and the Square Kilometer Array-Low, which will have an order of magnitude more collecting area than HERA and an instrument design well-suited to imaging.

We will next briefly describe the sensitivity of these arrays to the cosmological signal. We will see that the signal-to-noise per pixel is very small, except on the largest scales.  Thus, imaging is not possible: measurements focus on statistical quantities like the power spectrum.  For the sake of brevity, we will skip most of the relevant derivations and refer the reader to more thorough sources (e.g., \cite{furl06-review, parsons12, liu14a}).
 
\subsection{Sensitivity to the 21-cm Signal} \label{sens}
 
The sensitivity of a radio telescope depends on the competition between the signal strength ($T_b$) and the noise, which we express as $T_{\rm sys}$, the {\it system temperature},  defined as the temperature of a matched resistor input to an ideal noise-free receiver that produces the same noise power level as measured at the actual receiver's output.  The system temperature includes contributions from the telescope, the receiver system, and the sky; the latter dominates in our case.  For a single dish, noise fluctuations $\Delta T^N$ decline with increased bandwidth $\Delta \nu$ and integration time $ t_{\rm int}$ according to the radiometer equation,
\begin{equation}
\Delta T^N = \kappa_c\frac{T_{\rm sys}}{\sqrt{ \Delta \nu \, t_{\rm
int}}} \approx\frac{T_{\rm sys}}{\sqrt{ \Delta \nu \, t_{\rm int}}},
\label{eq:radiometer}
\end{equation}
where $\kappa_c\ge 1$ is an efficiency factor accounting for the
details of the signal detection scheme; for simplicity we will set
$\kappa_c=1$, which is a reasonable approximation for the telescopes
discussed here. 

Using equation (\ref{eq:tsky}) with $T_{\rm sys} \approx T_{\rm sky}$ to estimate the telescope noise $\Delta T^N$ for a single-dish measurement of an unresolved source, we find
\begin{equation}
\Delta T^N|_{\rm sd} \approx 0.6 \mkel \ \left( \frac{1+z}{10}
\right)^{2.6} \, \left( \frac{{\rm MHz}}{\Delta \nu} \, \frac{100
\hr}{t_{\rm int}} \right)^{1/2}.
\label{eq:single-dish}
\end{equation}
The mean 21-cm signal has $T_0 \sim 20 \mkel$; thus, single dish
telescopes can easily reach the sensitivity necessary to detect the
global 21-cm background. In this regime, the challenge is instead to
separate the slowly varying cosmological signal from the
foregrounds. 

However, at meter wavelengths the angular resolution of single dipoles or dishes is generally extremely poor, so mapping or statistical experiments require interferometers.  In that case,
\begin{equation}
\Delta T^N|_{\rm int} \sim 2 \mkel \ \left( \frac{A_{\rm tot}}{10^5 \sqm} \right) \, \left( \frac{10'}{\theta_D} \right)^2 \, \left( \frac{1+z}{10} \right)^{4.6} \, \left( \frac{{\rm MHz}}{\Delta \nu} \,
\frac{100 \hr}{t_{\rm int}} \right)^{1/2}.
\label{eq:if-sens}
\end{equation}
The angular resolution scale of $\theta_D\sim 10'$ and the frequency
resolution scale of $\Delta\nu\sim 1~{\rm MHz}$ correspond to $\sim
20$~comoving 
Mpc.\footnote{More precisely, a bandwidth $\Delta\nu$ corresponds
to a comoving distance $\sim 1.8~{\rm Mpc}(\Delta\nu/0.1~{\rm
MHz})[(1+z)/10]^{1/2}$, while an angular scale $\theta_D$ corresponds
to $2.7 (\theta_D/1')[(1+z)/10]^{0.2}$~Mpc.} The current generation of
telescopes have $A_{\rm tot} \la 10^5 \sqm$, so imaging (i.e., mapping pixels with a signal-to-noise much greater than unity) will only be possible on large scales that exceed the typical sizes of bubbles
during most of reionization.  Thus near-term imaging experiments focus primarily on giant H~II regions, such as those generated by extremely luminous quasars, where the contrast between the large ionized bubble and the background IGM is largest \cite{wyithe04-qso, lidz07}.

Although equation~(\ref{eq:if-sens}) provides a simple estimate of an
interferometer's sensitivity, the rate at which
interferometers sample different physical scales actually depends on the antenna distribution, making the sensitivity a function of angular and frequency scales.  Thus,
equation~(\ref{eq:if-sens}) only provides a rough guide.  A more precise estimate of the sample variance and thermal errors on the power spectrum is \cite{mcquinn06-param, furl07-cross}
\begin{equation}
\delta P_{21}(\bk_i) = P_{21}(\bk_i) + {T_{\rm sys}^2 \over B t_{\rm int}} \, {D^2 \Delta D \over n(k_\perp)} \left( {\lambda^2 \over A_e} \right)^2.
\label{eq:pkerror}
\end{equation}
This expression requires some unpacking. First, we assume we are observing a mode with wavenumber ${\bf k}_i$ in an observation that spans $t_{\rm int}$ \emph{total} time. We separate the components of that mode into those on the plane of the sky ($k_\perp$) and those along the line of sight $k_\parallel$. The former are affected by the angular resolution of the telescope, while the latter depend on the frequency resolution. They have fundamentally different systematics that are crucial for foreground removal strategies, as we will see below.  In any case, only a fraction of the total integration time is spent observing any given mode; the factor $n(k_\perp)$ incorporates that, is it is the number density of baselines observing a given wave mode, normalized to the total number of baselines in the instrument. $D$ is the comoving distance to the observed volume, and $\Delta D$ is the line-of-sight depth of that volume (which depends on the bandwidth of the observation). Finally, $A_e$ is the area of a single antenna in the array; the factor $A_e/\lambda^2$ is the angular resolution of the telescope in Fourier space (where the power spectrum lives).  In equation~(\ref{eq:pkerror}), the first term represents the sample variance within the observed volume, and the second is thermal noise.

In a real measurement, we will bin closely-spaced Fourier modes together to estimate the power spectrum. The number of Fourier cells in each power spectrum bin,
which depends on the Fourier-space resolution of the instrument, is
\begin{equation}
N_c(k) \approx 2 \pi \, k^2 \, \Delta k \, \Delta \mu \times \left[ \frac{V_{\rm surv}}{(2 \pi)^3} \right],
\label{eq:nc-annulus}
\end{equation}
where the last term represents the Fourier space resolution and we have grouped Fourier cells into annuli of constant $(k,\,\mu)$ (following the discussion of redshift-space distortions above).  The
total errors from all estimates within a bin simply add in quadrature.

To make estimates we must determine the baseline distribution $n(k_\perp)$ as well as the sampling density (equation~\ref{eq:nc-annulus} for a measurement in annuli).  These two quantities depend sensitively on the design of the experiment.  To develop intuition, it is therefore useful to consider the simple thermal noise-dominated case \cite{morales05}.  Substituting for $N_c$ in equation~(\ref{eq:pkerror}), ignoring the first term (which is equivalent to working on small scales), and assuming bin sizes $\Delta k \propto k$ and constant $\Delta \mu$, we find
\begin{equation}
\delta P_{\Delta T} \propto A_e^{-3/2} \, B^{-1/2} \, \left[ \frac{1}{k^{3/2} \, n(k,\, \mu)} \right] \, \left( \frac{T_{\rm sys}^2}{t_{\rm int}} \right).
\label{eq:sens-scale}
\end{equation}
This implies \cite{morales05}:

\begin{enumerate}

\item $\delta P_{21} \propto t_{\rm int}^{-1}$, because the power spectrum depends on the square of the intensity.  

\item We can increase the collecting area in two ways.  One is
to add antennae while holding the dish area $A_e$ constant.  Recall
that $n(k,\, \mu)$ is normalized to the total number of baselines $N_B
\propto N_a^2$: thus, adding antennae of a fixed size decreases the
errors by the total collecting area squared.  (Of course, the number
of correlations needed also increases by the same factor, so this
strategy is costly in terms of computing.)  The other method is to make each antenna
larger but hold their total number fixed.  In this case, the total
number of baselines, and hence $n(k,\, \mu)$, remains constant, so
$\delta P_{\Delta T} \propto A_e^{-3/2}$.  Increasing the collecting
area in this way is not as efficient because it decreases the total
field of view of the instrument, which is set by the field of view of each antenna.

\item Adding bandwidth increases the sensitivity relatively
slowly: $\delta P_{\Delta T} \propto B^{-1/2}$, because it adds new
volume along the line of sight without affecting the noise on any
given measurement.  Of course, one must be wary of adding too much
bandwidth because of systematics (especially foregrounds).

\item As a function of scale $k$, $\delta P_{\Delta T}
\propto k^{-3/2} \,n(k,\, \mu)^{-1}$.  The first factor comes from the
increasing (logarithmic) volume of each annulus as $k$ increases.  But
in realistic circumstances the sensitivity actually decreases toward
smaller scales because of $n$.  This is most obvious if we consider a
map at a single frequency.  In that case, high-$k$ modes correspond to
small angular separations or large baselines; for a fixed collecting
area the array must therefore be more dilute and the sensitivity per
pixel decreases as in equation (\ref{eq:if-sens}).  In the (simple but
unrealistic) case of uniform $uv$ coverage, the error on a measurement
of the angular power spectrum increases like $\theta_D^{-2}$ for a
fixed collecting area.

Fortunately, the three-dimensional nature of the true 21-cm signal
moderates this rapid decline toward smaller scales: even a single dish
can measure structure along the line of sight on small physical
scales.  Mathematically, because $n(k,\, \mu)=n(k_\perp)$, each
baseline can image arbitrarily large $k_\parallel$, at least in
principle.  For an interferometer, this implies that short baselines
still contribute to measuring large-$k$ modes.  Thus, provided that
they have good frequency resolution, compact arrays are surprisingly
effective at measuring small-scale power .  There is one important
caveat: if short wavelength modes are only sampled along the frequency
axis, we can only measure modes with $\mu^2 \approx 1$.  Thus we
recover little, if any, information on the $\mu$ dependence of the
redshift-space distortions.  Studying this aspect of the signal
\emph{does} require baselines able to measure the short transverse
modes with $\mu^2 \approx 0$.

\end{enumerate}

Figure~\ref{fig:sphere-sens} shows the expected errors (including only thermal noise and cosmic variance, not systematics) on the spherically-averaged power spectrum for an expanded version of the MWA (with double the current number of antenna elements), PAPER, and HERA.  (For the latter, see also Fig.~\ref{fig:foreground-strategy} below for a more detailed estimate.)  Current instruments may detect the signal, if they reach their design limits, but in most plausible scenarios it will be a tentative detection at best.  Next-generation experiments will be necessary for precision constraints.

Because the sky noise increases rapidly with redshift (see eq.~\ref{eq:tsky}), the first generation of experiments lose sensitivity at $z \sim 11$--12.  However, relatively modest expansions, like the HERA telescope now under development, can be optimized to make high-precision measurements out to $z \sim 15$.  One trick has become popular in the community for such purposes, which is to use redundant baselines \cite{parsons12}. Given a fixed array design, the best sensitivity for a statistical measurement is achieved when the signal-to-noise \emph{per mode} is unity.  Because a typical 21-cm machine is well below this threshold, it is useful to use \emph{redundant baselines} -- in which many different antenna pairs measure the same modes on the sky -- to move closer to the optimal measurement. Thus, rather than distribute baselines randomly to achieve maximal $uv$ coverage, as in a traditional interferometer, many 21-cm instruments follow very regular spacings.  Under some circumstances, this can also greatly accelerate the correlation computations by using a Fast Fourier Transform or one of its cousins \cite{tegmark10}, as pioneered by the OmniScope \cite{zheng13}.

Significantly larger instruments will be necessary to measure anisotropies (such as redshift-space distortions) in the spin-flip background, because they require separate measurements of power along the line of sight and across the plane of the sky \cite{mcquinn06-param}.  Although compact experiments can achieve high sensitivity on small scales by measuring frequency structure, they will not achieve the requisite sensitivity on the plane of the sky to measure angular fluctuations. The Square Kilometer Array may have the collecting area and angular resolution to perform such tests.

\subsection{Systematics and Foregrounds} \label{foregrounds}

Raw sensitivity to the cosmic 21-cm background is hard enough to achieve, but an additional difficulty is separating that signal from the many (and much, much brighter) astrophysical foregrounds, especially synchrotron emission from our Galaxy.  Conceptually, the way to separate these is straightforward (e.g., \cite{zald04, morales05_foregrounds}): most known foregrounds have very smooth spectra, while the cosmological signal varies rapidly as any given line of sight passes through density and temperature fluctuations and/or ionized bubbles.  If one imagines transforming the data into an ``image cube" (with observed frequency a proxy for radial distance), one ought to be able to fit a smooth function to each line of sight, subtract that smooth component, and be left with a measurement of the rapidly-varying cosmological component (plus any rapidly varying foregrounds).

\begin{figure}[t]
\centerline{\includegraphics[width=9cm]{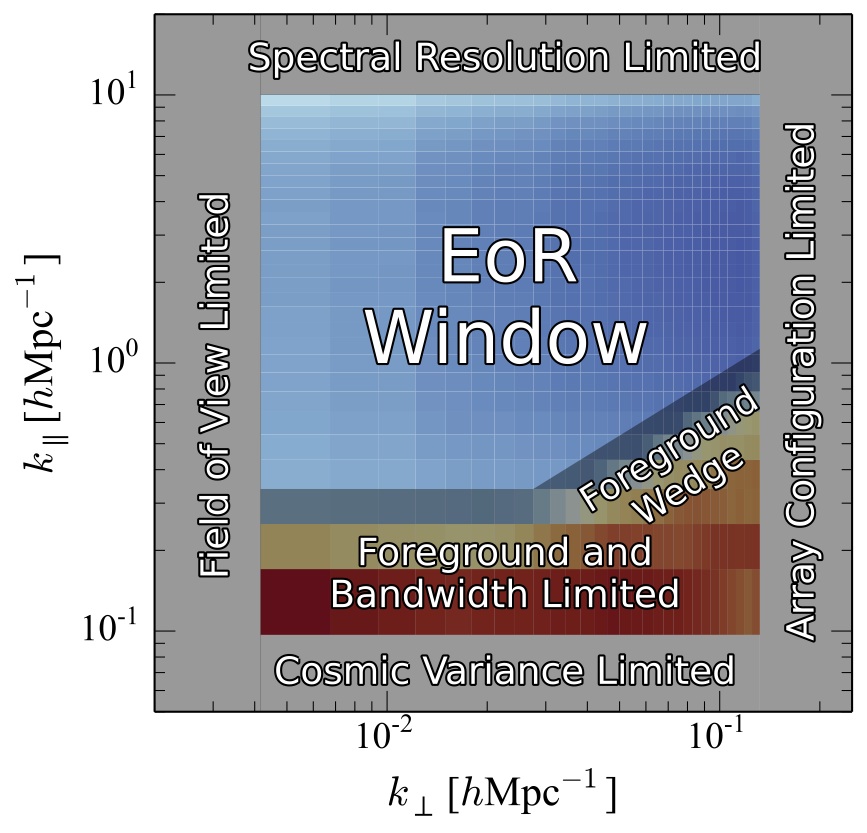}}
\caption{A schematic of the EoR window in the cylindrical $k_\perp$-$k_\parallel$ Fourier plane. At the smallest $k_\perp$, errors increase because of the instrumentÕs finite field of view. The largest $k_\perp$ that can be probed is determined by the interferometer's longest baseline. Similarly, the measurement's spectral resolution limits the sensitivity at large $k_\parallel$. In principle, cosmic variance determines the smallest measurable $k_\parallel$, but in practice the finite bandwidth and foreground contamination are more restrictive. As one moves towards larger $k_\perp$, however, the foregrounds leak out to higher $k_\parallel$ in a characteristic shape known as the \emph{foreground wedge}. The remainder of the Fourier plane has errors dominated by thermal noise, allowing (with a large collecting area or a long integration time) a clean measurement of the power spectrum in this \emph{EoR window}. From \cite{liu14a}.}
\label{fig:eor-window}
\end{figure}

Figure~\ref{fig:eor-window} illustrates how this limits the signal in Fourier space:\footnote{Here we think of the data as a ``Fourier cube," with $k_\parallel$ (derived by transforming the frequency) standing in for the radial direction.} small $k_\parallel$ modes have very large errors because they get confused with variations in the foregrounds.  Other more prosaic issues also limit the range of $k$-space to be sampled, including the finite bandwidth of the instrument, the field of view, the longest baselines in the array (which determine the largest measurable $k_\perp$), and the spectral resolution.  

As a mathematical exercise applied directly to, e.g., simulation boxes, these strategies work extremely well: they do indeed impose a minimum $k_\parallel$ but do not contaminate the data in the remainder of the measured region \cite{mcquinn06-param, liu11}. However, the practical details of this foreground removal are quite challenging (e.g., \cite{vedantham12}), and there has not yet been a successful application of these removal strategies to real-world data. The simplest challenge to understand is the intrinsic chromaticity of the interferometer: each baseline measures $k_\perp \propto D/\lambda$, where $D$ is the physical distance between the interferometer elements. Thus the instrument response changes across the measurement band, introducing spurious frequency-dependent features from foregrounds.

Chromatic effects such as these manifest themselves along a ``wedge" in Fourier space at large $k_\perp$ and small $k_\parallel$ \cite{datta10, morales12, parsons12b}, as shown schematically in Figure~\ref{fig:eor-window} (see \cite{pober13} and \cite{dillon14} for examples with real data from PAPER and the MWA).  Crucially, at least to the limits of current MWA and PAPER data, the ``extra" foreground contamination \emph{only} appears in this wedge.  This can be understood most simply by analyzing the data on a ``per-baseline" basis, by Fourier transforming each of the interferometer's baselines separately along the frequency axis: it arises from the relative delay between signals entering the two antennae in each baseline from different directions. This understanding (which can be described analytically \cite{liu14a,liu14b} can be used to minimize the impact of the wedge in new experiments (a key motivation in the design of HERA, for example).

In the near-term, the community's focus has thus shifted from \emph{foreground removal} to \emph{foreground avoidance}: if the problematic area is the wedge in Fourier space illustrated by Figure~\ref{fig:eor-window}, the simplest approach is to simply ignore data in that region and work inside the \emph{EoR window} that remains uncontaminated.  In practice, of course, there will be some residual contamination even here, from such factors as baseline gridding \cite{hazelton13}, ionospheric refraction and reflection, and polarized foreground leakage (which has strong frequency dependence due to Faraday rotation). But the PAPER team has already demonstrated a four order-of-magnitude reduction in foreground contamination \cite{parsons14}, as illustrated by Figure~\ref{fig:paper-wedge}.

\begin{figure}[t]
\sidecaption[t]
\includegraphics[width=7cm]{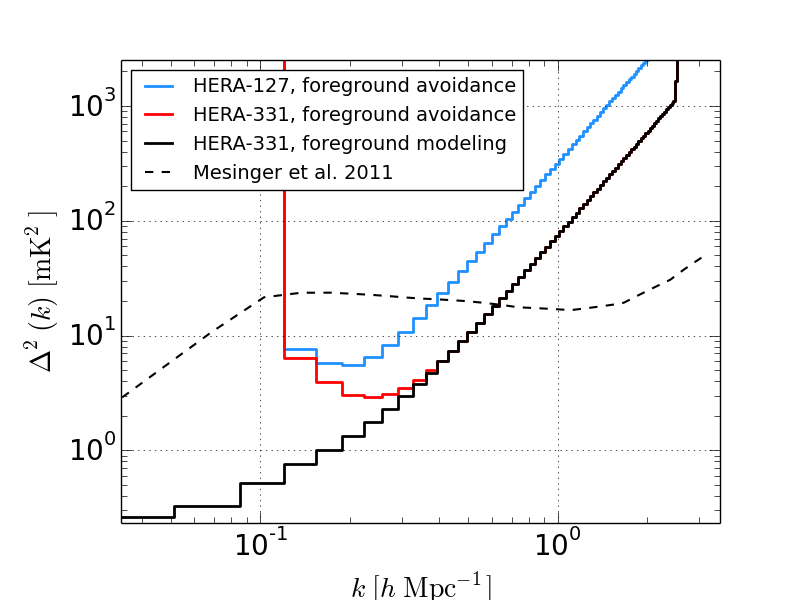}
\caption{HERA's power-spectrum sensitivity (solid lines) relative to a fiducial ionization model (dotted line; $x_{\rm HI} = 0.37$ at  $z = 9$; from \cite{mesinger11}). The blue curve represents a partial array with 127 dishes.  The red curve represents the full array, with 331 dishes.  Both of these curves assume that measurements cannot be made within the foreground wedge.  The black curve assumes that foregrounds can be removed in that area of the Fourier cube. Courtesy A. Liu, J. Pober, and J. Dillon.}
\label{fig:foreground-strategy}
\end{figure}

Still, the foreground wedge itself contains a great deal of astrophysical information: we can see this in Figure~\ref{fig:foreground-strategy}, which shows the sensitivity of the full HERA array assuming that foregrounds can be modeled accurately (black curve) versus a scenario where data within the wedge must be ignored (red curve). Thus the avoidance strategy is by no means optimal, and removal algorithms are still an active field of research. Many strategies appeal to careful calibration and image-based model subtraction \cite{zald04,liu09, bowman09, harker09}.  Others rely on optimal estimators and/or decorrelation to expand the window within which foregrounds are suppressed \cite{liu14b}. The challenge is greatly eased in the imaging regime, where the ability to isolate ionized bubbles allows one also to isolate the foregrounds at several discrete frequencies along each line of sight \cite{petrovic11}.  These strategies are useful not only for future analyses that hope to work \emph{within} the wedge but also for experiments that focus on foreground avoidance, because they also minimize leakage from the wedge into the EoR window.

Working within this area will ultimately be important not only for statistical detections but also for imaging campaigns, a key focus of the SKA telescope, as a full reconstruction of the signal with an interferometer requires dense coverage throughout the Fourier cube. The SKA should have a large enough collecting area for such efforts, which are especially important in relating the large-scale morphology of the reionization process to source maps produced with other telescopes.

\begin{acknowledgement}
I thank Adrian Liu for helpful comments on the manuscript.
\end{acknowledgement}

\bibliographystyle{unsrt}
\bibliography{Ref_composite}

\end{document}